\documentclass[lettersize,journal]{IEEEtran}
\usepackage{amsmath,amsfonts}
\usepackage{algorithmicx}
\usepackage{array}
\usepackage[caption=false,font=normalsize,labelfont=sf,textfont=sf]{subfig}
\usepackage{textcomp}
\usepackage{stfloats}
\usepackage{url}
\usepackage{verbatim}
\usepackage{graphicx}
\usepackage{cite}
\usepackage{multirow}
\usepackage{picinpar}
\usepackage{amsmath}
\usepackage[latin1]{inputenc}
\usepackage{colortbl}
\usepackage{soul}
\usepackage{multirow}
\usepackage{pifont}
\usepackage{color}
\usepackage{alltt}
\usepackage[hidelinks]{hyperref}
\usepackage{enumerate}
\usepackage{siunitx}
\usepackage{breakurl}
\usepackage{epstopdf}
\usepackage{pbox}
\usepackage{bm}
\usepackage{booktabs}
\usepackage{color}
\usepackage{amssymb}
\usepackage{makecell}
\usepackage[ruled,vlined]{algorithm2e}

\begin{document}

\title{Optimization of Flying Ad Hoc Network Topology \\and Collaborative Path Planning for Multiple UAVs}

\author{Ming He,
        Peizhao Wang,
        Haihua Chen,~\IEEEmembership{Member,~IEEE,}
        Bin Sun,
	  Hongpeng Wang~\IEEEmembership{Member,~IEEE}
\thanks{Manuscript received 12 November 2023; revised xx Month 2024.}
\thanks{This work was supported in part by the National Natural Science Foundation of China under Grants  61973173 and 62373201, in part by the Technology Research and Development Program of Tianjin under Grant 18ZXZNGX00340. (Corresponding author: H. Chen)

M. He, P.-Z. Wang, H.-H. Chen and B. Sun are with the College of Electronic Information and Optical Engineering, Nankai University, Tianjin, 300071, China. (e-mail: heming@nankai.edu.cn; wangpz@mail.nankai.edu.cn; hhchen@nankai.edu.cn; 2120220403@mail.nankai.edu.cn).

H.-P. Wang is with the College of Artificial Intelligence, Nankai University, Tianjin, 300071, China, and the Laboratory of Science and Technology on Integrated Logistics Support, National University of Defense Technology, Changsha 410073, China. (e-mail: hpwang@nankai.edu.cn).}
}

\markboth{Journal of \LaTeX\ Class Files,~Vol.~14, No.~8, August~2021}%
{Shell \MakeLowercase{\textit{et al.}}: A Sample Article Using IEEEtran.cls for IEEE Journals}

\maketitle

\begin{abstract}
Multiple unmanned aerial vehicles (UAVs) play a vital role in monitoring and data collection in wide area environments with harsh conditions. In most scenarios, issues such as real-time data retrieval and real-time UAV positioning are often disregarded, essentially neglecting the communication constraints. In this paper, we comprehensively address both the coverage of the target area and the data transmission capabilities of the flying \emph{ad hoc} network (FANET). The data throughput of the network is therefore maximized by optimizing the network topology and the UAV trajectories. The resultant optimization problem is effectively solved by the proposed reinforcement learning-based trajectory planning (RL-TP) algorithm and the convex-based topology optimization (C-TOP) algorithm sequentially. The RL-TP optimizes the UAV paths while considering the constraints of FANET. The C-TOP maximizes the data throughput of the network while simultaneously constraining the neighbors and transmit powers of the UAVs, which is shown to be a convex problem that can be efficiently solved in polynomial time. Simulations and field experimental results show that the proposed optimization strategy can effectively plan the UAV trajectories and significantly improve the data throughput of the FANET over the adaptive local minimum spanning tree (A-LMST) and cyclic pruning-assisted power optimization (CPAPO) methods.
\end{abstract}



\section{Introduction}
\IEEEPARstart{M}{onitoring} tasks are generally demanding in forest, desert, alpine tundra and other wide-area environments, where infrastructure and human resources are scarce. However, relying solely on manpower to complete these tasks can be challenging and time consuming. Unmanned aerial vehicles (UAVs) are therefore introduced as a substitute for humans, and multiple UAVs compose a flying \emph{ad hoc} network (FANET) to cover a wide area. FANET has attracted significant interest and found many applications in electric power inspection, security, urban mapping, and so on.

Most traditional planning problems for multiple UAVs focused on the efficiency of task completion while overlooking real-time data sharing and inter-UAV cooperation. In the event that one or more UAVs become disabled during the task, ensuring data integrity and promptly recovering impaired UAVs becomes challenging, which hinders the practical applications of multi-UAV systems. In light of this trend, traditional task allocation methods for multiple UAVs without real-time data links fail to meet these requirements, which highlights the necessity for the joint optimization of task allocation and FANET topology.

\begin{figure}[!t]
\centering
\includegraphics[width=3.5in]{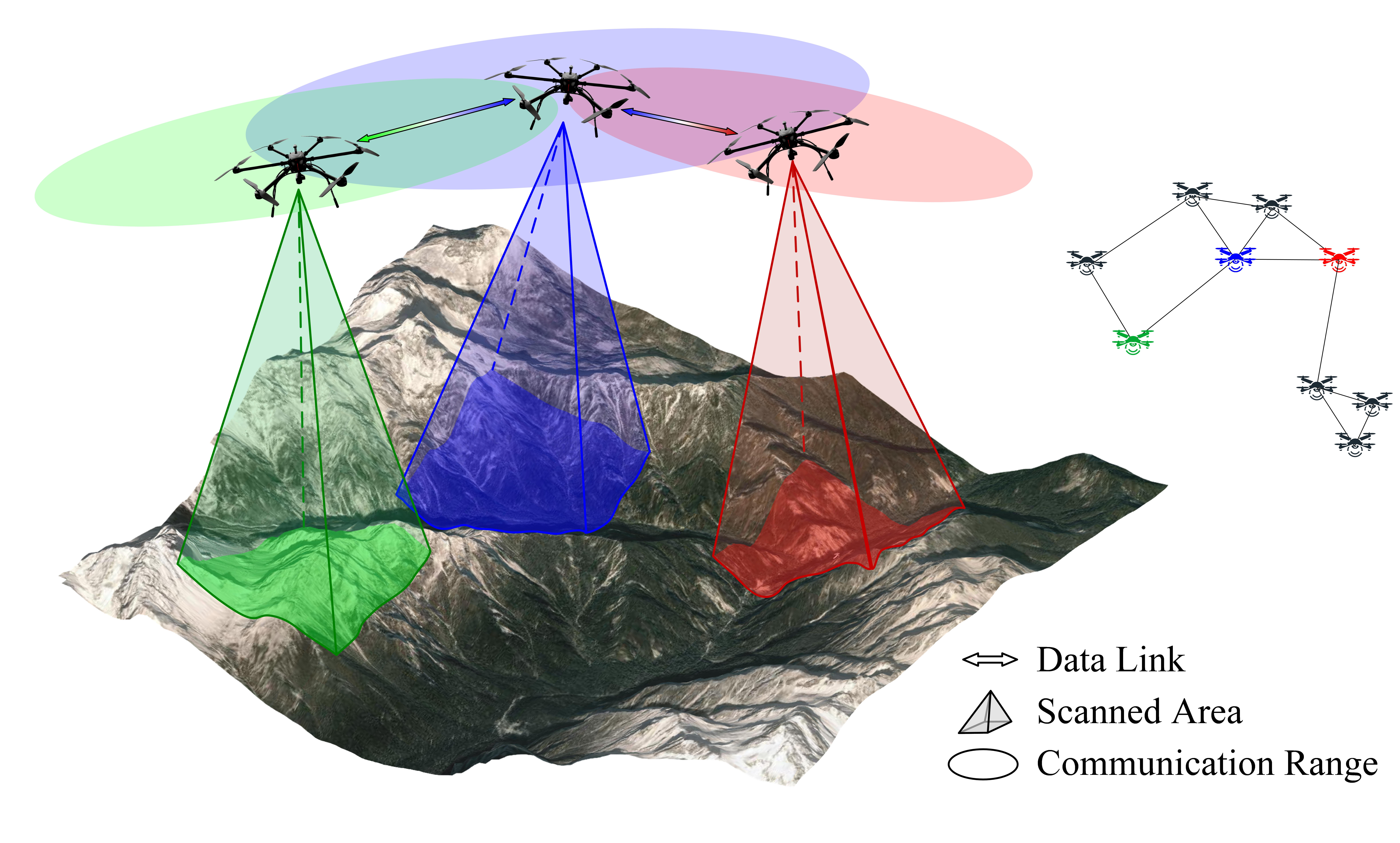}
\caption{Illustration of the scenario for cooperative multi-UAV networks. The UAVs form a FANET and are planned to traverse all the waypoints to visually cover a wide-area. The FANET is fully connected in each time slot.}
\label{fig1}
\vspace{-0.5cm}
\end{figure}

This paper addresses a scenario as depicted in Fig. \ref{fig1}, where multiple \textcolor{black}{multirotor} UAVs equipped with cameras are deployed to visually scan a rugged mountain. Real-time monitoring of the UAVs' status and their collected data is required throughout the entire process. It is therefore necessary to maintain at least one reachable data link between any pair of UAVs to facilitate the transmission of real-time images and possible environment parameters. Optimal UAV paths and FANET topology can improve the efficiency and fulfillment rate of the process, but the non-convexity and tight coupling between variables of this multivariate optimization problem make it difficult to obtain a closed-form solution, which poses a significant challenge in maximizing the data throughput of a FANET while simultaneously satisfying the coverage requirements and communication constraints.


In this paper, the multivariate optimization problem is formulated as maximizing the total data throughput of the entire network over the task duration, subject to the constraints of topology and visual coverage. The variables in the resultant problem are decoupled and the problem is solved efficiently using the algorithm of reinforcement learning based trajectory planning (RL-TP) and convex-based algorithm of topology optimization (C-TOP). The main contributions of this paper are outlined as follows.

\begin{enumerate}
        \item To guarantee the data communication link in the autonomous decision-making systems, we propose a new framework of joint FANET topology optimization (JFTO) for communication and path planning of multiple UAVs. Different from the most state-of-the-art frameworks, it considers the practical situation that the communication range of the UAVs is limited.
        
        \item To achieve an optimal trade-off between network robustness and spatial reuse, we propose a novel algorithm that simultaneously optimizes the number of neighbors and transmit powers in FANETs, which enables the programmability of the autonomous decision-making systems.
        
        \item To improve the quality of data links, we propose the C-TOP algorithm, which transforms the optimization problem of the transmit powers of the UAVs into a convex problem. The C-TOP algorithm guarantees the optimal solution can be found in polynomial time, which enables the online realization of the autonomous decision-making systems.
\end{enumerate}

The rest of this paper is organized as follows. The related work of this paper is reviewed in Section II. In Section III, the system is modeled as an optimization problem with multiple constraints. In Section IV, a three-step approach is proposed to solve the optimization problem, which consists of a path planning approach, a topology optimization algorithm, and a convex optimization-based algorithm. Experimental results are provided in Section V to illustrate the performance of the proposed approach. Finally, Section VI concludes the paper.

\section{Related Work}
In recent years, there has been a growing research interest in multi-UAV-based systems, prompting extensive efforts to enhance the quality of task completion across diverse scenarios. In most scenarios deploying multiple UAVs for monitoring tasks, the focus was primarily on addressing path planning or tracking issues. Consequently, this field has seen the development of numerous exemplary algorithms, broadly categorized into two main groups.

The first category is represented by classical methods such as dynamic programming \cite{DP} and derivative correlation \cite{2022DCOC}. The system models considered in these algorithms are relatively simple, with high computation complexity and easy to fall into local optima. The other category is referred to as meta-heuristic algorithms. In recent years, greedy algorithm \textcolor{black}{\cite{greedy}}, genetic algorithm (GA) \textcolor{black}{\cite{GA1}}, ant colony optimization algorithm (ACO) \cite{CVRP} and particle swarm optimization method (PSO) \textcolor{black}{\cite{PSO}} are also widely used to solve the optimization problem of path planning in FANET. These methods speed up the convergence rate and alleviate sinking into the local optima by imitating biological habits. In addition, the problem of the collaborative path planning is approximately solved using convex optimization, which substantially reduces the computation complexity when the number of UAVs is large \cite{cvxpp}. In the last five years, artificial intelligence (AI) based technologies have also been proposed to find the optimal path for UAVs \cite{Attention}. However, the AI based methods have high training costs when the destined map is large.

Most of the aforementioned cooperative planning algorithms for multiple UAVs neglect the communication topology, resulting in the challenge of recovering impaired UAVs at a short time and ensuring data integrity when some UAVs fail. Topology control algorithms for static wireless networks are dominated by graph theory, such as minimum spanning tree (MST) \cite{MST}, relative neighbor graph (RNG) \cite{RNG} and Yao graph \cite{Yao}. Most of the graph-theory based topology control methods optimize the overall connectivity of the network without considering the spatial reuse and the random link disconnection. The $k$-connected network topologies are then proposed to counter the failure of $k$ nodes, including the cone-based topology control (CBTC) \cite{CBTC} and fault-tolerant topology control (FTTC) \cite{FTTC}. In multi-UAV systems, due to the fast mobility of the UAVs, the FANETs are usually time varying. In \cite{Mobility1}, the mobility model of nodes \textcolor{black}{is} taken into account to optimize the topology, but the application scenario has limitations. \cite{Mobility3} \textcolor{black}{provides} virtual force-based mobility control (VFMC) to ensure stable bi-connectivity. In order to consider both the system tasks and the network performance, the nodes of the FANET \textcolor{black}{are} designed to move around the optimal point in \cite{FT2}. However, it is difficult for the UAVs to visually scan the target area. \cite{Mobility4} \textcolor{black}{proposes} an algorithm that cooperatively control the multiple UAVs in the scenario of disaster search and rescue, which considers the communications between the ground searchers and the UAVs. However, the optimization of the network topology was not included.

Recently, due to the advancements in communication technology, there has been increasing research on integrating communication technology into various fields. An optimization problem of communication in energy-constrained and UAV-assisted relay networks \textcolor{black}{is} addressed in \cite{JO1}. The problem \textcolor{black}{is} decoupled into three sub-problems and an iterative optimization approach \textcolor{black}{is} employed, using successive convex approximation (SCA) to obtain the final solution. A near-optimal solution \textcolor{black}{is} found by the hybrid intelligent algorithm that combines a stochastic simulation and improved GA \cite{JO2}. The algorithm \textcolor{black}{is} proposed for airborne radar networks (ARNs) in the presence of multi-uncertainty. A joint communication motion planning (JCMP) method for minimizing the total energy of communication and motion in the scenarios consisting of single- and multi-sensing robots \textcolor{black}{is} introduced in \cite{JO3}. To solve the problem that the multi-UAV systems are susceptible to interference, an alternating control strategy of power and mobility based on interactive reward region (IRR) \textcolor{black}{is} proposed in \cite{JO4}. In the communication scenario involving active eavesdropping UAVs, an algorithm \textcolor{black}{is} proposed to maximize the effective eavesdropping rate \cite{JO5}, which \textcolor{black}{optimizes} the jamming power and path planning sequentially. Although the network performance is improved by optimizing the location and transmit power of the UAVs, the network topology is not directly optimized. In addition, the robustness of the network is rarely considered, which is necessary as the UAVs are vulnerable in the wild areas.

In this paper, a joint optimization framework involving UAV trajectory and network topology is proposed for three-dimensional (3D) visual coverage scenarios using multiple UAVs. The trajectories of the UAVs are cooperatively planned and the network topology is optimized as well. In addition, the robustness of the network is considered in the optimization framework by including the constraints of the number of the neighbors. The optimization problem is decoupled into three sub-problems, which can be solved using the proposed RL-TP and C-TOP algorithms.



\section{Problem Formulation}
\subsection{Assumptions}
\begin{figure}[!t]
\centering
\includegraphics[height= 1.2in]{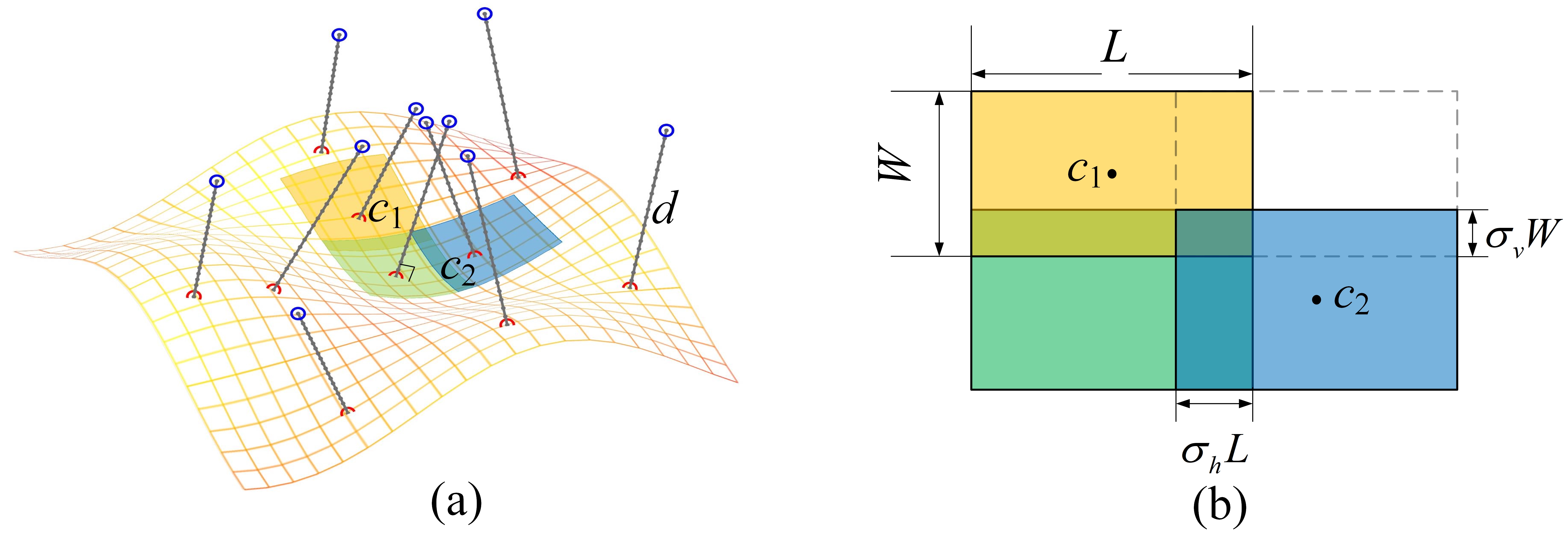}
\caption{\textcolor{black}{Schematic of geometric constraints for photographic alignment. (a) Equidistant and frontal alignment constraints. (b) Overlapping requirements for discrete waypoints.}}
\label{fig_of_waypoint}
\vspace*{-1.5em}
\end{figure}

Consider a scenario where multiple UAVs equipped with cameras are tasked with visually covering a wide 3D area of a known model. \textcolor{black}{The waypoints for multirotors are determined by using the digital elevation model (DEM), with the constraints that the cameras are equidistant and frontal photogrammetric \cite{2022DEM}. As shown in Fig.~\ref{fig_of_waypoint}(a), the cameras are adjusted to be perpendicular to the smooth 3D terrain surface $\mathcal{D}$ and maintain an equal distance from the target to ensure a consistent image resolution. The aerial waypoints are labeled based on a grid-like partitioning of the target area, ensuring full coverage with minimal overlap.  For a camera with a coverage area of $L \times W$, let $c_1$ and $c_2$ represent two adjacent waypoints. The coverage areas of these adjacent waypoints are required to partially overlap, with the overlapping region having a length of $\sigma_h L$ and a width of $\sigma_v W$, where $\sigma_h$ and $\sigma_v$ are two constants such that $0<\sigma_h < 1$ and $0<\sigma_v<1$, as illustrated in Fig.~\ref{fig_of_waypoint}(b). This overlap is designed to ensure continuity between adjacent images, which helps improve the accuracy and quality of image stitching and 3D reconstruction.}

The objective is to optimize the FANET topology and determine the paths for all UAVs that satisfy the system requirements. The following assumptions are made regarding the system:

\begin{enumerate}
    \item Path planning: each aerial waypoint is visited once and only once, and the UAVs are not required to return to the starting points when all the waypoints are visited.
    \item Data link: UAVs transmit data using orthogonal channels with a half-duplex antenna, and the channel between any two UAVs is reciprocal.
	\end{enumerate}

\subsection{Definitions}\label{sec-def}
According to the requirements of image overlap for 3D reconstruction and the control characteristics of \textcolor{black}{multirotor} UAVs, the problem of visual coverage of a bounded 3D region $\mathcal{D}\subset\mathbb{R}^3$ by multiple UAVs can be transformed into the task of traversing a set of aerial waypoints \cite{2022DEM}. Due to the requirement of data transmission, the traversal problem is formulated into an optimization problem that maximizes the data throughput of the network subject to the constraints of the FANET topology and the trajectory length.

In addition, the number and locations of the starting points are chosen according to the distribution of the waypoints to be traversed. The sets of the starting points and the waypoints are denoted by $\mathcal{S}$ and $\mathcal{W}$, respectively.

Let $\mathcal{A}=\{a|a=1,2,\cdots,A\}$ be the set of the heterogeneous UAVs, where $A$ is the number of available UAVs. The optimized trajectory of UAV $a$ is denoted by $J_a=(j_{0,a}, \allowbreak j_{1,a},\dots, j_{M_a,a})$, where $j_{0,a} \in \mathcal{S} $ is the starting point, $j_{m,a} \in \mathcal{W} $, $m=1,2,\cdots, M_a$, is the waypoint traversed by UAV $a$, and $M_a$ is the number of waypoints visited by UAV $a$. \textcolor{black}{An example of $J_a$ is $J_a=(2,6,10,5)$, where the starting waypoint is $2$, $M_a=3$, and UAV $a$ sequentially traverses the waypoints $6$, $10$, and $5$.} The trajectories of all the UAVs constitute the set $\mathcal{J}$, i.e. $\mathcal{J}=\{J_a|a=1,2,\cdots,A\}$.

A 3D matrix ${\bf{E}} \triangleq [e_{i,j,a}]$ is introduced to denote the paths that are visited by the UAVs, where $[e_{i,j,a}]$ stands for the notation that the $(i,j,a)$-th element of the matrix is $e_{i,j,a}$. Matrix ${\bf{E}}$ has a dimension of $(S+W)\times (S+W) \times A$, where $S = |\mathcal{S}|$, $W=|\mathcal{W}|$, and $|\cdot|$ denotes the number of a set. The value of $e_{i,j,a}$ is set to $1$ if UAV $a$ covers the path from waypoint $i$ to waypoint $j$. Otherwise $e_{i,j,a}=0$. \textcolor{black}{For the example $J_a=(2,6,10,5)$, the elements $e_{2,6,a}=1$, $e_{6,10,a}=1$, $e_{10,5,a}=1$, and all other elements in the $a$-th page of matrix ${\bf{E}}$ are zero.} Let $l_{i,j}$ denote the path length between the waypoints $i$ and $j$. The $(S+W)\times (S+W)$ matrix ${\bf{L}} \triangleq [l_{i,j}]$ is introduced. The lengths of the trajectories of all the UAVs can therefore be determined by matrices ${\bf{L}}$ and ${\bf{E}}$ together.

The paths and the time varying topology of the FANET are optimized subject to the constraints of system energy and duration. The discrete time topology is therefore introduced to model the time varying network. We assume that the maximal allowed duration of UAV $a$ is $T_{a, \rm max}$, and the maximal $T_{a, \rm max}$ is divided into $N$ time slots ${t_n} \buildrel \Delta \over = \frac{n}{N}{\rm{max}}_a({T_{a,{\rm{max}}}})$, $n=0,1,\cdots,N$. For notation simplicity, set $\mathcal{N} \triangleq \{n|n=0,1,\cdots,N\}$ is defined.

\begin{figure}[!t]
\centering
\includegraphics[height=1.2in]{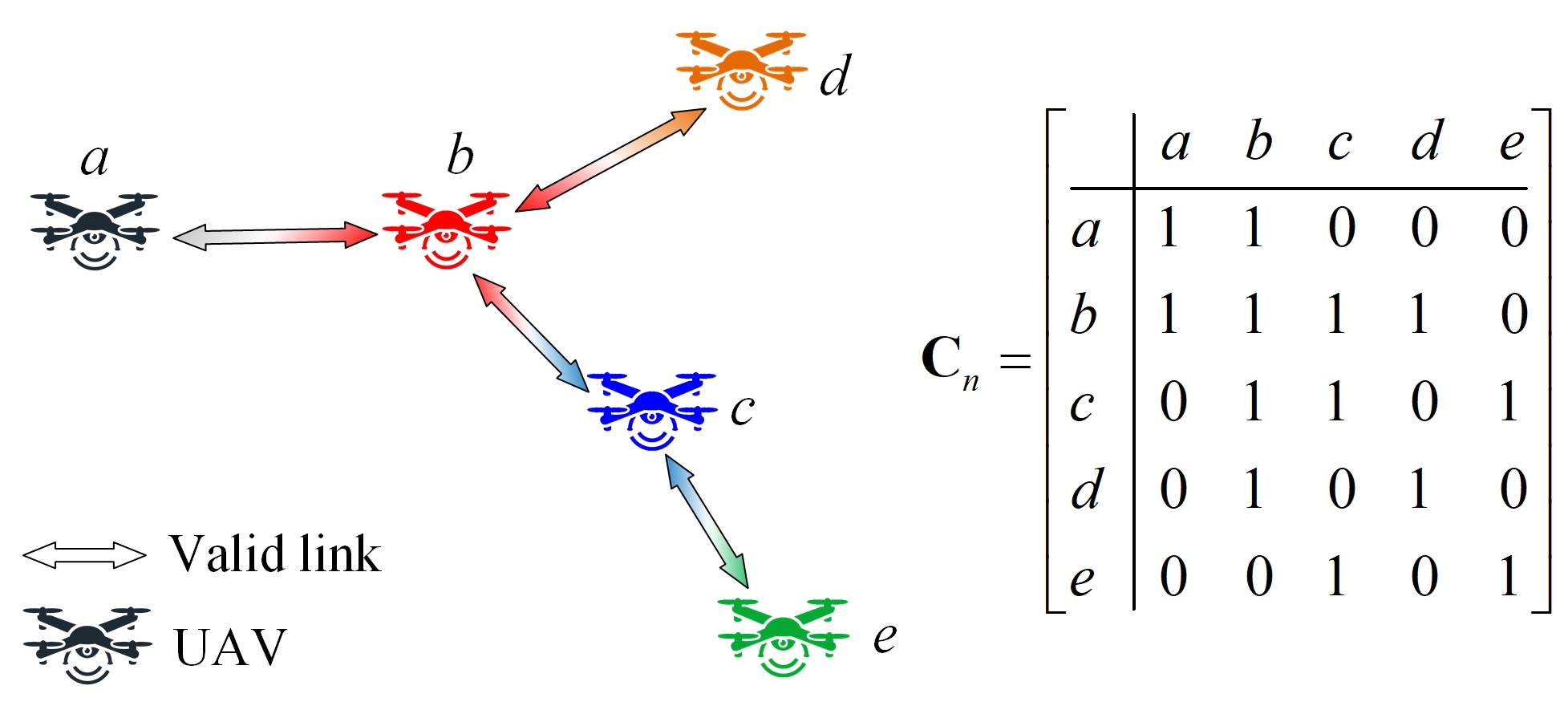}
\caption{\textcolor{black}{Example of network topology and the corresponding reachability matrix.}}
\label{fig_topology_eg}
\vspace*{-0.3em}
\end{figure}

The topology of the network in each time slot $t_n$ is modeled as an undirected graph $\mathcal{G}_n \triangleq (\mathcal{A},{\bf C}_n)$, where ${\bf C}_n \triangleq [c_{a,b,n}]$ is the reachability matrix that indicates the connection state of the UAVs in the $n$-th time slot \textcolor{black}{\cite{C_matrix}}. Matrix ${\bf C}_n$ has a dimension of $A\times A$ and $c_{a,b,n} \in \{0,1\}$. \textcolor{black}{Fig.~\ref{fig_topology_eg} shows an example of the network topology and its corresponding reachability matrix for $5$ UAVs in the $n$-th time slot. 
}

Let the transmit power of UAV $a$ in the $n$-th time slot be $p_{a,n}$. The received power of UAV $b$ is written as $p_{b,n}^{\rm R} = p_{a,n} h_{a,b,n}$, where $h_{a,b,n}$ is the path loss from UAV $a$ to UAV $b$ in the $n$-th time slot.

A valid data link between UAVs $a$ and $b$ is established or they are neighbors if $p_{b,n}^{\rm R}$ is higher than the sensitivity threshold $\gamma$ \textcolor{black}{\cite{threshold}}, i.e.
\begin{eqnarray}
  \vspace*{-0.5em}
  c_{a,b,n}&=&\left\{
  \begin{tabular}{ll}
    $1$, & $p_{b,n}^{\rm R} \geq \gamma$  \\
    $0$, & $p_{b,n}^{\rm R} < \gamma$
  \end{tabular} \right. \label{cab}
\end{eqnarray}
The elements on the diagonal of ${\bf C}_n$ are set to 1 because each UAV is always considered to have a valid 
link with itself.

The path loss $h_{a,b,n}$ is determined by the distance between UAVs and the environment around. It is assumed in this paper that the path between the transmitter and the receiver is unobstructed since the altitudes of UAVs are usually high enough. Therefore, the free space propagation model is used \textcolor{black}{\cite{path_loss}}, the path loss of which can be written as
\begin{eqnarray}
  \vspace*{-0.5em}
  h_{a,b,n}&=& \frac{\mu_f}{d_{a,b,n}^2} \label{hab}
\end{eqnarray}
where $\mu_f$ is a constant that depends on the characteristics of the transceivers. $d_{a,b,n}$ denotes the spatial distance between UAVs $a$ and $b$ at time slot $n$, which is determined by the trajectories in $\mathcal{J}$. Due to the assumption that the channel between any two UAVs is reciprocal, ${\bf C}_n$ is symmetric. 

The number of neighbors of a node in the FANET is an important parameter that determines the network connectivity, \textcolor{black}{which corresponds to the node degree in graph theory\cite{node_degree}}. UAV $a$ and UAV $b$ are neighbors of each other at time slot $n$ if $c_{a,b,n} = 1$. Let $k_{a,n}$ be the number of neighbors of UAV $a$ in time slot $n$. We have
\begin{eqnarray}
  k_{a,n} &=& \sum_{b=1,b\neq a}^{A} c_{a,b,n} \label{ka}
\end{eqnarray}
It can be seen from (\ref{cab})-(\ref{ka}) that the connectivity of the network can be controlled by varying the transmit power of the UAVs.

Throughput determines the data transmission capacity of a wireless network, and it is therefore maximized to optimize the network topology. The throughput from UAV $a$ to UAV $b$ in the $n$-th time slot can be written as \textcolor{black}{\cite{CDS_throughput}}
\begin{eqnarray}
    {R_{a,b,n}} (p_{a,n}) &=& B\log_2 ( 1 + \frac{p_{a,n} h_{a,b,n}} {\sigma^2 B} ) \label{Rab}
\end{eqnarray}
where $B$ is the bandwidth of the wireless channel, $\sigma^2$ is the variance of the receive noise. 


\vspace*{-1em}
\subsection{Problem Statement}
The ultimate goal of this paper is to find the optimal paths ${\mathcal J}$, the transmit powers ${\mathcal P}$, and the topologies ${\mathcal G}$ of the UAVs by maximizing the throughput ${\mathcal T}$ of the FANET. The constraints on the path planning, the topology, and the power define a solution set ${\mathcal C}$, which are shown in (\ref{P0once1}) - (\ref{P0emax}). The problem is solved using a three-step algorithm. Use the definitions defined in Section \ref{sec-def}, the optimization problem can be written as
\begin{subequations}\label{P0}
\allowdisplaybreaks
    \begin{align}
        \mathop {\max }\limits_{{\mathcal J},{\mathcal P},{\mathcal G}} \; &{\mathcal T} \triangleq  \sum\limits_{n = 1}^N {\sum\limits_{b = a + 1}^A {\sum\limits_{a = 1}^A {{c_{a,b,n}}} } } {R_{a,b,n}}({p_{a,n}}) \label{P0obj}\\
        {\rm s.t.} \;  & \sum_{a=1}^{A} \sum_{i=S+1}^{S+W} e_{i,j,a}=1, \;\; j = S+1,\cdots,S+W \label{P0once1}\\
        & \sum_{a=1}^{A} \sum_{j=S+1}^{S+W} e_{i,j,a}=1, \;\; i = S+1,\cdots,S+W \label{P0once2}\\
        & {d_{a,b,n}} \ge {d_{\min }}, \hspace*{5.4em} a,b \in {\mathcal A}, n \in {\mathcal N}\label{P0avoid}\\
        & \sum_{a=1}^{A} \sum_{i=1}^{S+W}\sum_{j=1}^{S+W} e_{i,j,a} l_{i,j} \le L_{\rm max}\label{P0length}\\
        & k_{a,n} \geq K_{\rm min} , \hspace*{5.6em} a \in \mathcal{A}, n \in \mathcal{N}  \label{P0kmin}\\
        &k_{a,n} - K_{\rm min} \leq \delta, \hspace*{4.0em} a \in \mathcal{A}, n \in \mathcal{N} \label{P0kmax}\\
        &{\bf C}_n^{H} > {\bf 0}, \hspace*{7.4em} n \in \mathcal{N}\label{P0_fullC} \\
        &0 \leq p_{a,n} \leq P_{a, \rm max}, \hspace*{3.9em} a \in \mathcal{A}, n \in \mathcal{N}     \label{P0pmax}\\
        &\sum_{n=0}^{N} \frac{T_{\rm max}}{N} p_{a,n}\leq E_{a, \rm max}, \hspace*{1.4em} a \in \mathcal{A}\label{P0emax}
    \end{align}
\end{subequations}
where $\mathcal{G} \triangleq \{\mathcal{G}_n|n=0,1,\cdots,N\}$ is the set of the topologies in all the time slots, $\mathcal{P} \triangleq \{p_{a,n}|a=1,2,\cdots,A, n=0,1,\cdots,N\}$ is the set of the transmit powers of all the UAVs in all the time slots. It can be seen from (\ref{P0obj}) that the objective of the problem is to maximize the total data throughput and the solution to the problem is the optimal trajectories, topologies, and transmit powers of the UAVs.

The constraints in (\ref{P0once1}) and (\ref{P0once2}) together restrict each waypoint to be traversed by one and only one UAV \textcolor{black}{to avoid the search area being scanned repeatedly.} Each of the starting points, however, can have multiple UAVs. The constraint (\ref{P0avoid}) imposes a restriction on the minimum distance between any pair of UAVs, which is the anti-collision constraint. The constraint on the total length of the trajectories is shown in (\ref{P0length}), where $L_{\rm max}$ is the maximal allowed length of the paths.

In (\ref{P0kmin}), the connectivity of the network topology is specified by requiring the minimum number of neighbors of each node to be $K_{\rm min}$, which improves the robustness of the network and avoids high network latency. For example, if $K_{\rm min}$ is set to a value of 2, then each UAV has at least two links in each time slot. In such case, the network keeps connected even if one of its neighbors is disabled. 
In addition, the maximum number of neighbors of the UAVs is also constrained in (\ref{P0kmax}), which restricts the number of ``extra'' neighbors in a range specified by the integer parameter $\delta$. Order the neighbors of UAV $a$ by the distance from it, then the ``extra'' neighbors are referred to those that have an order number larger than $K_{\rm min}$. \textcolor{black}{Constraint (\ref{P0kmax}) is considered in the optimization to guarantee the spatial reuse. The spatial reuse denotes the communication capacity for multiple UAVs to concurrently utilize the same frequency spectrum without incurring intolerable interference within a specified communication radius. An increase in $k_{a,n}$ leads to a larger number of neighboring UAVs. When the number of communication neighbors is higher, the chances of communication interference also increase, resulting in reduced spatial reuse of the communication channels.}
Therefore, the lower the value of $k_{a,n}$ is, the higher the spatial reuse is, because fewer neighbors are impacted by the data transmission of the nodes \textcolor{black}{\cite{space_reuse}}. 

To guarantee that the resultant network is globally connected, all the elements of the $H$-th power of ${\bf C}_n$ are required to be non-zero in constraint (\ref{P0_fullC}), where $H \triangleq A-K_{\rm min}$, and ${\bf 0}$ denotes a matrix whose elements are all zeros \textcolor{black}{\cite{reachable}}.

Due to the limited energy of the device, the transmit power and the total energy of data transmission of the UAVs are constrained as shown in (\ref{P0pmax}) and (\ref{P0emax}), respectively, where $P_{a, \rm max}$ and $E_{a, \rm max}$ are the maximal allowed transmit power and energy, respectively. The constraint on the kinetic energy is not included in the problem as it is assumed that all the UAVs have enough energy for the traversing in $T_{a, \rm max}$.

\section{Proposed Joint FANET Topology Optimization}
It can be seen that the problem in (\ref{P0}) is equivalent to finding the optimal paths $\mathcal{J}$, optimal topologies $\mathcal{G}$, and optimal transmit powers $\mathcal{P}$ that maximize the network throughput. \textcolor{black}{This problem is NP-hard due to its nonlinear and non-convex nature \cite{CDS_throughput}}. The primary difficulty solving this problem lies in the strong coupling among the variables. In this paper, we first decouple the problem into two parts: path optimization, which focuses on finding the optimal paths for UAVs, and topology optimization, which focuses on determining the communication links between UAVs.

\begin{figure*}[!t]
\centering
\includegraphics[width=6.8in]{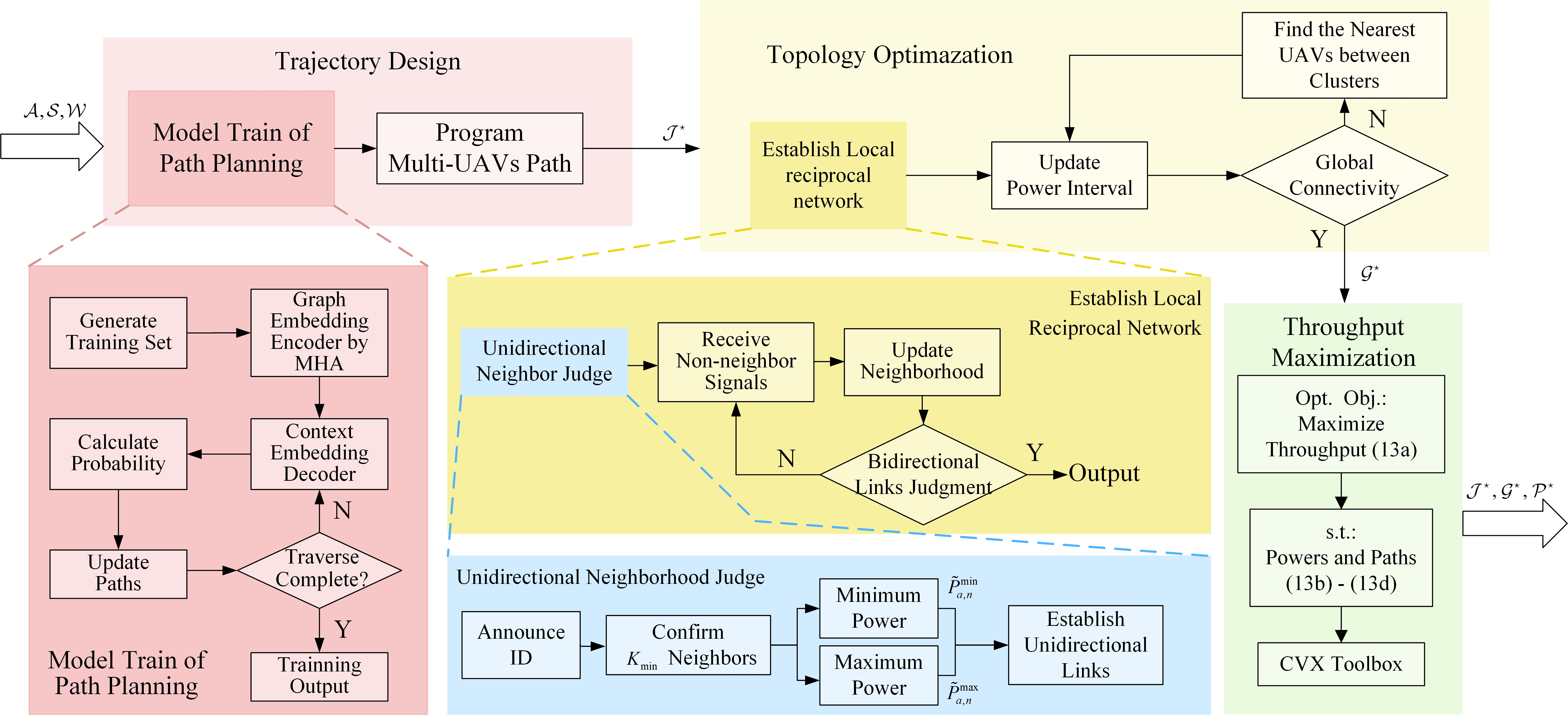}
\caption{System diagram. The system consists of three parts: path planning, topology optimization and maximization of network throughput.}
\label{system_diagram}
\vspace{-1em}
\end{figure*}

\textcolor{black}{During the phase of path optimization, the transmit powers are fixed at the maximal allowed value}, and the RL-TP algorithm is employed to find the optimal paths. In the phase of topology optimization, the paths are set as the optimal solution obtained in the phase of path optimization, and the focus then shifts to optimizing the communication topology between UAVs. Since the integer variable $c_{a,b,n}$ is a step function of the real-valued power $p_{b,n}^{\rm R}$, it is challenging to find the optimal solution for $\mathcal{G}$ and $\mathcal{P}$. To maximize the network throughput, this paper further decouples the topology and node power. Once the topology is determined, the overall system performance is optimized by adjusting the power levels of each node. In summary, this section proposes an efficient three-step algorithm to decouple constraint (\ref{P0kmin}), and the constraints of $\mathcal{G}$ and $\mathcal{P}$.


A diagram of the proposed algorithm is shown in Fig. \ref{system_diagram}. As shown in the figure, the data sets $\mathcal{A}$, $\mathcal{S}$, $\mathcal{W}$ are initialized at the beginning of the algorithm. The optimal trajectories of the UAVs are then determined by solving (\ref{P1}). The optimal trajectories $\mathcal{J}^{\star}$ are subsequently used to optimize the network topology, as presented in problem (\ref{P3}). The transmit powers of the UAVs in different time slots are then obtained by solving the problem in (\ref{P4}). The output of the system consists of the optimal trajectories, topologies, and transmit powers.

\subsection{Trajectory Design} \label{sec-trajectory}
The first step in decoupling the variables is to find the optimal paths traversing all the waypoints by assuming that the UAVs transmit signals with their maximal allowed transmit power ${\bf{P}}_{\rm max}$. By using (\ref{P0}), the decoupled problem can then be written as
\begin{subequations}\label{P1}
\allowdisplaybreaks
    \begin{align}
        \underset {\mathcal{J}, \mathcal{G}} {\min} & \sum_{a=1}^{A} \sum_{i=1}^{S+W}\sum_{j=1}^{S+W} e_{i,j,a} l_{i,j} \label{P1obj}\\
        {\rm s.t.} \;  & \eqref{P0once1}-\eqref{P0avoid}, \label{P1one}\\
        & \tilde t_{M_a,a}\le T_{a, \rm max}, \label{P1tmax}\\
        & \tilde k_{a,n} \geq K_{\rm min}, \hspace*{3.5em} a \in \mathcal{A}, n \in \mathcal{N}   \label{P1kmin}
    \end{align}
\end{subequations}
where $\tilde t_{M_a,a}$ represents the moment when UAV $a$ reaches its last waypoint $j_{M_a,a}$. Constraint \eqref{P1tmax} limits the maximum flight time of UAVs. $\tilde k_{a,n} = \sum_{b=1,b\neq a}^{A} \tilde c_{a,b,n}$ and $\tilde c_{a,b,n}$ is defined similarly to $c_{a,b,n}$ in (\ref{cab}), except that the transmit power of UAV $a$ is $p_{a,n}=P_{a, \rm max}$. The constraint that restricts the number of extra neighbors is not considered here for simplicity. However, the constraint in (\ref{P1kmin}) ensures the UAVs form topologies in which each node has at least $K_{\rm min}$ neighbors.

\textcolor{black}{It can be seen in (\ref{P1obj}) that problem (\ref{P1}) is formulated to directly minimize the total path length, thereby inherently satisfying the constraint specified in (\ref{P0length}). Additionally, problem (\ref{P1}) serves as a subproblem of (\ref{P0}), and its feasibility can be determined by comparing the optimal objective value of (\ref{P1}) with the requirement in (\ref{P0length}).} The optimal solution to problem (\ref{P1}) is a set of ordered waypoints. In fact, the problem reduces to the capacitated vehicle routing problem (CVRP) \cite{CVRP} if constraint (\ref{P1kmin}) is excluded and the UAVs take a round trip from the same point. CVRP is an NP-hard problem and has been approximately solved by various heuristic algorithms \cite{CVRPsolution}. However, the existing algorithms for CVRP cannot be applied straightforwardly to solving problem (\ref{P1}) since they are not generally constrained to have neighbors and other dynamic conditions. Recently, reinforcement learning-based methods have shown their effectiveness and efficiency in addressing CVRP-like problems. We therefore propose a modified RL-TP \cite{RLTP} to solve problem (\ref{P1}). Compared with the traditional RL-TP method, the proposed RL-TP is capable of solving the problem that includes the constraint in \eqref{P1kmin}, which is introduced to ensure that the topology of the time-varying trajectory satisfies the requirements of data communication.



\begin{figure*}[t]
\centering
\includegraphics[width=1\textwidth]{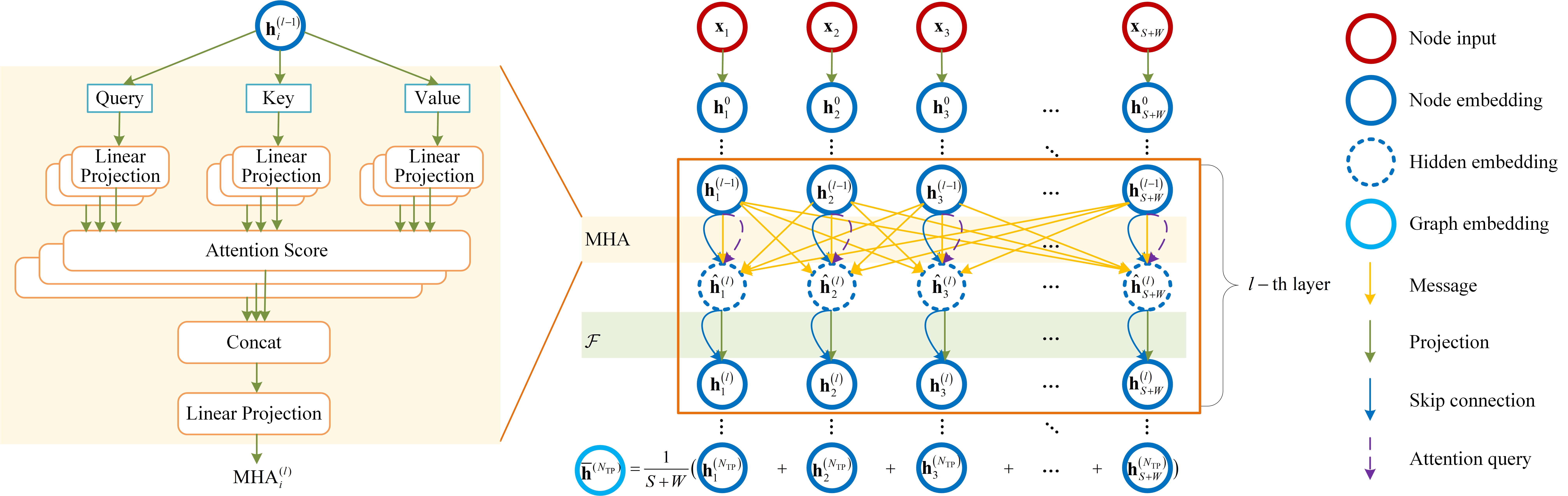}
\caption{\textcolor{black}{Structure of the encoder. Input nodes are embedded and processed through $N_{\rm TP}$ sequential layers. The graph embedding is the mean of the final node embeddings. Each MHA sublayer consists of $H$ attention heads, each with a dimension of $D \times 1$, for feature extraction.}}
\label{MHA_structure}
\end{figure*}

\textcolor{black}{As shown in the red part of Fig.~\ref{system_diagram}, the multi-head attention (MHA) mechanism is used for trajectory planning due to its ability to capture spatial relationships and long-range dependencies among waypoints \cite{Attention}. MHA allows the model to attend to multiple parts of the input sequence concurrently, providing a richer representation of the environment and effectively addressing complex dependencies, such as communication topology constraints. By using multiple attention heads, the model can explore different trajectory options in parallel, leading to more informed decisions and improved path planning performance, particularly in environments where obstacles and connectivity constraints influence optimal path selection.
}

The MHA-based RL-TP includes an encoder and a decoder. In the encoder, the node coordinates of the graph are input to the transformer network to generate both node and graph embeddings, which are then input to the decoder. In the decoder, the sampling distribution for selecting the next point is calculated using a single-head attention layer, which leverages the graph information obtained from the encoder. The probability of coordinate selection is then reinforced using a greedy rollout baseline, which effectively accelerates the solving process and convergence.

The modified RL-TP includes the following three steps.
\subsubsection{Encoding}
As shown in Fig.~\ref{MHA_structure}, the encoder consists of $N_{\rm TP}$ sequential attention layers, each of which has an MHA sublayer and a fully connected feed-forward (FCFF) sublayer. The input features include the number and coordinates of the waypoints, $\mathcal{S}$ and $\mathcal{W}$. Denote the 3D coordinate of waypoint $i$ as ${{\bf{x}}_i} = [{x_i},{y_i},{z_i}]^T$. The initial node embeddings are obtained by a linear projection
\begin{equation}
    {\bf{h}}_i^{(0)} = \left\{ {
    \begin{array}{*{20}{ll}}
        {{\bf{W}}_\mathcal{S}{{\bf{x}}_i} + {\bf{b}}_\mathcal{S}}&{i = 1,\cdots ,S}\\
        {{\bf{W}}_\mathcal{W}{{\bf{x}}_i} + {\bf{b}}_\mathcal{W}}&{i = S + 1,\cdots ,S + W}
    \end{array}} \right.
\end{equation}
where ${{\bf{W}}_{\cal S}}$, ${{\bf{b}}_{\cal S}}$, ${{\bf{W}}_{\cal W}}$, ${{\bf{b}}_{\cal W}}$ are trainable parameters.
The initial node embeddings are then processed through the attention layers. 

In the $l$-th layer, the output of the MHA sublayer for the $i$-th node can be expressed as 
\begin{equation}
    \begin{aligned}
        {\rm{MHA}}_i^{(l)}( &{{\bf{h}}_1^{(l-1)}, \cdots ,{\bf{h}}_{S + W}^{(l-1)}} ) \\
        & \hspace*{-3.5em} = {{\bf{W}}_O^{(l)}} {\rm{Concat(}} {\bf a}_{i,1}^{(l)}{\rm{,}} {\bf a}_{i,2}^{(l)}{\rm{,}} \cdots {\rm{,}} {\bf a}_{i,H}^{(l)}{\rm{)}}, l=1,2,\cdots, N_{\rm TP}
    \end{aligned}
\end{equation}
where ${{\bf{W}}_O^{(l)}}$ is the trainable output transformation matrix, $\rm{Concat}(\cdot)$ denotes the operation that concatenates multiple features into a single unified representation, $H$ is the number of attention heads,
\begin{equation}
    \begin{aligned}
        \!\!&{\bf a}_{i,h}^{(l)}= \\
        \!\!& {\rm{softmax}}\left( {\frac{{{{\left( {{\bf{W}}^{(l)}_{Q,h}{\bf{h}}_i^{(l-1)}} \right)}^T}{\bf{W}}^{(l)}_{K,h}{\bf{h}}_i^{(l-1)}}}{{\sqrt {{d_k}} }}} \right){\bf{W}}^{(l)}_{V,h}{\bf{h}}_i^{(l-1)} \\
        \!\!&\hspace*{5em} h = 1,2,\cdots, H
    \end{aligned}
\end{equation}
is of dimension $D\times 1$, and query matrix ${{\bf{W}}^{(l)}_{Q,h}}$, key matrix ${{\bf{W}}^{(l)}_{K,h}}$, and value matrix ${\bf{W}}^{(l)}_{V,h}$ are trainable parameters.

The output of the MHA sublayer is batch-normalized before input to the FCFF sublayer
\begin{equation}\label{h_hat}
    \hspace*{-1em} {{{\bf{\hat h}}}_i^{(l)}} = {\rm{ B}}{{\rm{N}}^{(l)}}\left( {{\bf{h}}_i^{(l - 1)} + {\rm{MHA}}_i^{(l)} \left( {{\bf{h}}_1^{(l - 1)}, \cdots ,{\bf{h}}_{S+W}^{(l - 1)}} \right)} \right)
\end{equation}
where ${\rm BN}^{(l)}(\cdot)$ is the operation of batch normalization (BN) \cite{BN}. The output of the $l$-th attention layer can be written as
\begin{equation}\label{hl}
    {\bf{h}}_i^{(l)} = {\rm{ B}}{{\rm{N}}^{(l)}}\left( {{{{\bf{\hat h}}}_i^{(l)}} + {{\mathcal{F}}^{(l)}}\left( {{{{\bf{\hat h}}}_i^{(l)}}} \right)} \right)
\end{equation}
where ${{\mathcal{F}}^{(l)}}( {{{{\bf{\hat h}}}_i^{(l)}}} )$ is the output of the FCFF sublayer \cite{FF}.

The output of the final layer, i.e. the node embeddings ${\bf{h}}_i^{(N_{\rm TP})}$, is averaged to obtain the graph embedding
\begin{equation}\label{final_EBD}
    {{{\bf{\bar h}}}^{(N_{\rm TP})}} = \frac{1}{{S + W}}\sum\limits_{i = 1}^{S + W} {{\bf{h}}_i^{(N_{\rm TP})}}
\end{equation}
Both the node and graph embeddings are provided as inputs to the decoder.

\subsubsection{Decoding}
\textcolor{black}{
The decoder derives the stochastic policy ${g_\theta }({\boldsymbol{\pi}}|s) = \prod_{j=1}^{A+W} {{g_\theta }({\pi _j}|s, \pi_{1},\pi_{2}, \cdots, \pi_{j - 1})}$ guided by the node and graph embeddings, where $\boldsymbol{\pi}=[\pi_1,\pi_2,\cdots,\pi_{A+W}]$ records the UAV paths, that is, $\boldsymbol{\pi} = \text{Concat}(J_1,J_2,\cdots,J_A)$, $\theta$ denotes all the trainable parameters in the encoder-decoder attention model, such as ${{\bf{W}}^{(l)}_{Q,h}}$, ${{\bf{W}}^{(l)}_{K,h}}$, and ${\bf{W}}^{(l)}_{V,h}$, $s$ is an instance of the path planning problem.}

The stochastic policy ${g_\theta }({\boldsymbol{\pi}}|s)$ and the UAV paths $\boldsymbol{\pi}$ are determined sequentially in $(A+W)$ steps. Let ${\bf{\Gamma }}^{(j)}\triangleq  \{ i|i \in {\mathcal S} \cup {\mathcal W},i \notin \{ {\pi _1}, \cdots ,{\pi _{j - 1}}\} \} $ denote the set of unallocated waypoints at the $j$-th step. The stochastic policy at the $j$-th step can then be written as
\begin{equation}\label{pro}
    {g_\theta }({\pi _j}|s, \pi_{1},\pi_{2}, \cdots, \pi_{j - 1}) = \max_{i \in {\bf{\Gamma }}^{(j)} } \frac{{{e^{{c_i}}}}}{{\sum\nolimits_{i'\in {\bf{\Gamma }}^{(j)}} {{e^{{c_{i'}}}}} }}
\end{equation}
where the logit $c_i$ is the raw score of each waypoint and is extracted by a single layer of head attention as
\begin{equation}
    {c_i} = \left\{ 
        {\begin{array}{*{20}{c}}
            {C \cdot {\rm{tanh}}\frac{{{{\left( {{{\bf{W}}_Q}{{\bf{h}}_{c}}} \right)}^T}{{\bf{W}}_K}{\bf{h}}_i^{(N_{\rm TP})}}}{{\sqrt {{d_k}} }}}&{{\rm{if}}\;i \in {\bf{\Gamma }}^{(j)}}\\
            { - \infty }&{{\rm{else}}}
        \end{array}} 
    \right.
\end{equation}
query vector ${{\bf{W}}_Q}$ and key vector ${{\bf{W}}_K}$ are trainable parameters, $C$ is a hyperparameter, and ${{\bf{h}}_{c}}$ is the context embedding generated by
\begin{equation} \label{hc}
    {{\bf{h}}_{c}} = \mathcal{F}\left( {{{{\bf{\bar h}}}^{({N_{{\rm{TP}}}})}}} \right) + \mathcal{F}\left( {{{\bf{h}}_{\pi }},{{\bf{h}}_{\Gamma}},{\bf{T}},{\bf{K}}} \right)
\end{equation}
In \eqref{hc}, ${{\bf{h}}_{\pi }}$ represents the concatenation of node embeddings for the $(j-1)$ already allocated waypoints, while ${{\bf{h}}_{\Gamma }}$ denotes the concatenation of node embeddings for the unallocated waypoints, ${\bf{T}}\triangleq [\tilde t_{M_a,a}]$ and ${\bf{K}}\triangleq [\tilde k_{a,n}]$ are system parameters, $\tilde t_{M_a,a}$ and $\tilde k_{a,n}$ are defined in problem (\ref{P1}).
\color{black}

\subsubsection{Training}
\textcolor{black}{To train the path planning model, we define the loss for the instance $s$ as $\mathcal{L}(\theta|s) = {E_{{g_\theta }(\boldsymbol{\pi} |s)}}\{L(\boldsymbol{\pi})\}$, representing the expectation of path length $L(\boldsymbol{\pi})$, which is used to update the parameters. The loss $\mathcal{L}(\theta|s)$ is optimized using gradient descent,} and an on-policy reinforcement learning method with a greedy rollout baseline is employed. The gradient of the loss is updated in each episode by
\textcolor{black}{\begin{equation}
    \begin{aligned}
        {\nabla _\theta }&\mathcal{L}(\theta |s) = {E_{{g_\theta }(\boldsymbol{\pi} |s)}}\\
        & \hspace*{-1.5em} \left\{ \!\! \left[ {\sum\limits_{j = 1}^{A+W} \!\! {{\nabla _\theta }\log {g_\theta }({\pi_j}|s,\pi_1,\cdots,\pi_{j-1})}}  \right] \!\!\! \left[ {\mathop \sum \limits_{j' = 1}^{A + W} \!\! {\tau ^{j'}}{r_{j'}} - b(s)} \right]\!\! \right\}
    \end{aligned}
\end{equation}}
\textcolor{black}{where $\tau$ is the discount factor of reward and is set to $1$ due to the stationary nature of the environment, implying that future rewards are considered equivalent in value to the immediate rewards,} $r_{j'}$ represents the immediate reward of each step in an episode, defined as the distance between two consecutive waypoints visited by a UAV, and $b(s)$ is the baseline of instance $s$. 

\textcolor{black}{With appropriately adjusted model parameters, the RL-TP method achieves the optimal solution by outputting the shortest feasible path $\boldsymbol{\pi}^\star$ on the test set upon convergence, while satisfying the optimization conditions in \eqref{P1}. The optimal path set $\mathcal{J}^\star$ is derived from $\boldsymbol{\pi}^\star$ by identifying the starting waypoints, where each segment between consecutive starting waypoints represents the optimal path for a single UAV.
}

\begin{figure}[!t]
\centering
\includegraphics[width=3in]{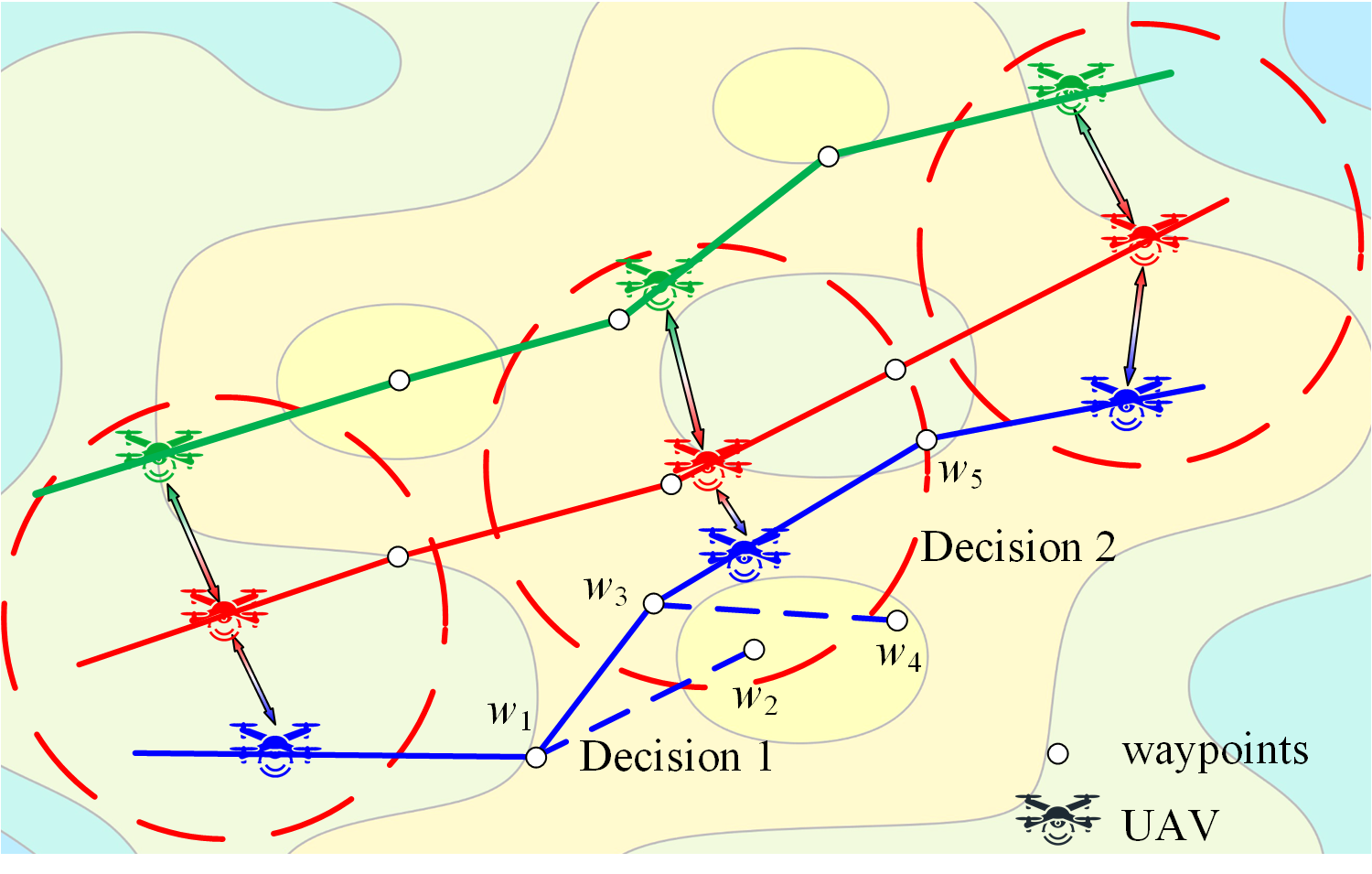}
\caption{\textcolor{black}{Example of path optimization. The dashed red circles denote the communication range of the UAV in some time slots. The solid and dashed segments stand for the optimized and unselected paths of the UAVs, respectively. The red UAV is connected to at least $K_{\rm min}$ neighbors in all the time slots.}}
\label{UAV_connected}
\end{figure}

\textcolor{black}{
The paths optimized by the RL-TP algorithm aims to find the shortest path while satisfying the topology constraints. As shown in Fig.~\ref{UAV_connected}, in the first decision of the blue UAV, path $(w_1,w_3)$ is selected, which concurrently fulfills the criteria for both the shortest path and topological constraints. However, there are scenarios that a shorter path is discarded as the path violates the topology constraints. For example, in the second decision of the blue UAV, the optimal path $(w_3,w_5)$ is longer than the infeasible path $(w_3,w_4)$.
}

\begin{algorithm}[!t]
\caption{RL-TP}\label{algorithm1_simple}
\KwIn{Number of epochs $E_{\rm TP}$, steps per epoch $M_{\rm TP}$, batch size $B_{\rm TP}$, significance $\alpha$}
\KwOut{The best model $\theta_{\rm BM}$, optimal paths $\mathcal{J}^\star$}

Initialize $\theta,\theta_{\rm BM} \leftarrow \theta$ 

\For{$e=1$ to $E_{\rm TP}$}
    {\For{$m=1$ to $M_{\rm TP}$}
        {\For{$b=1$ to $B_{\rm TP}$}
            {$s_b \leftarrow {\rm RandomInstance}()$
            
            $\boldsymbol{\pi}_b \leftarrow {\rm SampleRollout}({s_b},{g_\theta })$
            
            ${{\boldsymbol{\pi }}_{b,{\rm{BM}}}} \leftarrow {\rm{SampleRollout}}({s_b},{g_{{\theta _{{\rm{BM}}}}}})$}

        $\nabla {\mathcal L} \hspace*{-0.3em}\leftarrow \hspace*{-0.5em} \sum\nolimits_{b = 1}^{{B_{{\rm{TP}}}}} {(L({{\boldsymbol{\pi }}_b}) - L({{\boldsymbol{\pi }}_{b,{\rm BM}}})){\nabla _\theta }\log {g_\theta }({{\bf{\pi }}_b})}$
        
        $\theta \leftarrow {\rm Adam}(\theta,\nabla \mathcal{L})$
        
        \If{${\rm OneSidedPairedTTest}(g_\theta,g_{\theta_{\rm BM}}) < \alpha$}
            {$\theta_{\rm BM} \leftarrow \theta$} } }

{$\mathcal{J}^\star \leftarrow {g_{{\theta _{\rm BM}}}}(\boldsymbol{\pi}^\star |s)$}       
\end{algorithm}

\textcolor{black}{The computational complexity of the RL-TP is primarily driven by the MHA structure within the encoder. Each node embedding undergoes an MHA operation with a complexity of $\mathcal{O}(D_H^3)$, where $D_H\triangleq D\times H$ is the dimensionality of the node embedding. This operation is performed across $N_{\rm TP}$ layers for a total of $(S+W)$ waypoints. As a result, the overall computational complexity is $\mathcal{O}(D_H^3N_{\rm TP}(S + W))$, and the space complexity is $\mathcal{O}(D_H^2N)$, as the space is primarily used for storing node embeddings and intermediate values during the attention operations. The pseudocode for the RL-TP algorithm is provided in Alg.~\ref{algorithm1_simple}.}

\textcolor{black}{After obtaining the optimal path set $\mathcal{J}^\star$, we apply a segmented fifth-order B\'{e}zier curve to generate smooth trajectories for each UAV. The smoothing process is conducted using a hierarchical optimization approach that minimizes the total flight duration while ensuring continuity in position, velocity, and acceleration of each UAV, as detailed in \cite{2022DEM}. This method enhances path smoothness and improves the efficiency of flight search.}

\subsection{Topology Optimization}\label{sec-topology}
With the trajectories designed in Section \ref{sec-trajectory}, the problem in (\ref{P0}) is simplified to
\begin{subequations}\label{P2}
\allowdisplaybreaks
    \begin{align}
        \underset {\mathcal{P},\mathcal{G}} \max \;  &
        \sum _{n=1}^{N} \sum_{ b=1}^{A} \sum_{a=1}^{A} c_{a,b,n} R_{a,b,n} (p_{a,n}) \label{P2obj}\\
        {\rm s.t.} \;  & \eqref{P0kmin}-\eqref{P0emax}\\
        & J_a \in \mathcal{J}^\star, \hspace*{6.0em} a \in \mathcal{A} \label{P2J}
    \end{align}
\end{subequations}
where $\mathcal{J}^\star$ is the set of optimal trajectories designed in Section \ref{sec-trajectory}. \textcolor{black}{The problem in (\ref{P2}) remains NP-hard due to its combinatorial nature \cite{C_matrix}}. As can be seen from the constraint in (\ref{P0kmin}), the number of neighbors for each UAV is guaranteed to be at least $K_{\rm min}$, assuming that the transmit power is $P_{a, \rm max}$. However, the number of neighbors of each UAV should be restricted to ensure reasonable spatial reuse of the network. Therefore, the constraint in (\ref{P0kmax}) is imposed on the optimization problem. Instead of solving problem (\ref{P2}) directly, the optimal topology is first found before the optimal transmit power is determined. The optimization of the topology is formulated as
\begin{subequations}\label{P3}
\allowdisplaybreaks
    \begin{align}
        \underset {\tilde P_{a,n}^{\rm min},\tilde P_{a,n}^{\rm max}} \min \;  & \underset{a \in \mathcal{A}, n \in \mathcal{N}} \max{k_{a,n}^{\rm max}} & \label{P3obj}\\
        {\rm s.t.} \quad\;  & k_{a,n}^{\rm min} \geq K_{\rm min}, & a \in \mathcal{A}, n \in \mathcal{N}  \label{P3kmin}  \\
        &k_{a,n}^{\rm max} - K_{\rm min} \leq \delta, & a \in \mathcal{A}, n \in \mathcal{N}  \label{P3kmax}\\
        &{\bf C}_n^{H} > {\bf 0}, & n \in \mathcal{N} \hspace*{3.1em} \label{P3_fullC} \\
        &0 \leq \tilde P_{a,n}^{\rm max} \leq P_{a, \rm max}, &  a \in \mathcal{A}, n \in \mathcal{N}     \label{P3pmax}\\
        & J_a \in \mathcal{J}^\star, &  a \in \mathcal{A} \hspace*{3.2em} \label{P3J}
    \end{align}
\end{subequations}
where $\tilde P_{a,n}^{\rm min}$ and $\tilde P_{a,n}^{\rm max}$ are auxiliary variables representing the minimal and maximal allowed transmit powers of UAV $a$ in the $n$-th time slot, respectively, and $k_{a,n}^{\rm max}$ denotes the number of neighbors of UAV $a$ when $p_{a,n} = \tilde P_{a,n}^{\rm max}$. The flow of the C-TOP algorithm to solve this optimization problem is shown in the yellow and blue parts of Fig. \ref{system_diagram}, and the detailed steps are outlined as follows.

\subsubsection{Determine the minimum powers}\label{min_power}
The auxiliary variable $\tilde P_{a,n}^{\rm min}$ ensures that the number of neighbors of each UAV is greater than or equal to the required number $K_{\rm min}$. To satisfy the constraint in (\ref{P3kmin}), the received power of the nearest $K_{\rm min}$ UAVs must exceed the sensitivity threshold. Using (\ref{hab}), the value of $\tilde P_{a,n}^{\rm min}$ is determined by the distance between UAVs $a$ and $b$, which can be written as
\textcolor{black}{
\begin{eqnarray}
    \tilde P_{a,n}^{\rm min} &=& \frac{\gamma d_{a,b,n}^2} {\mu_f} \label{Pmin1}
\end{eqnarray}
The minimum transmit power $\tilde P_{a,n}^{\rm min}$ is smaller than the maximal allowed transmit power $P_{a, \rm max}$, as the UAV paths are optimized under the constraint of maintaining at least $K_{\min}$ neighbors.
}

\begin{figure}[!t]
\centering
\includegraphics[width=3.5in]{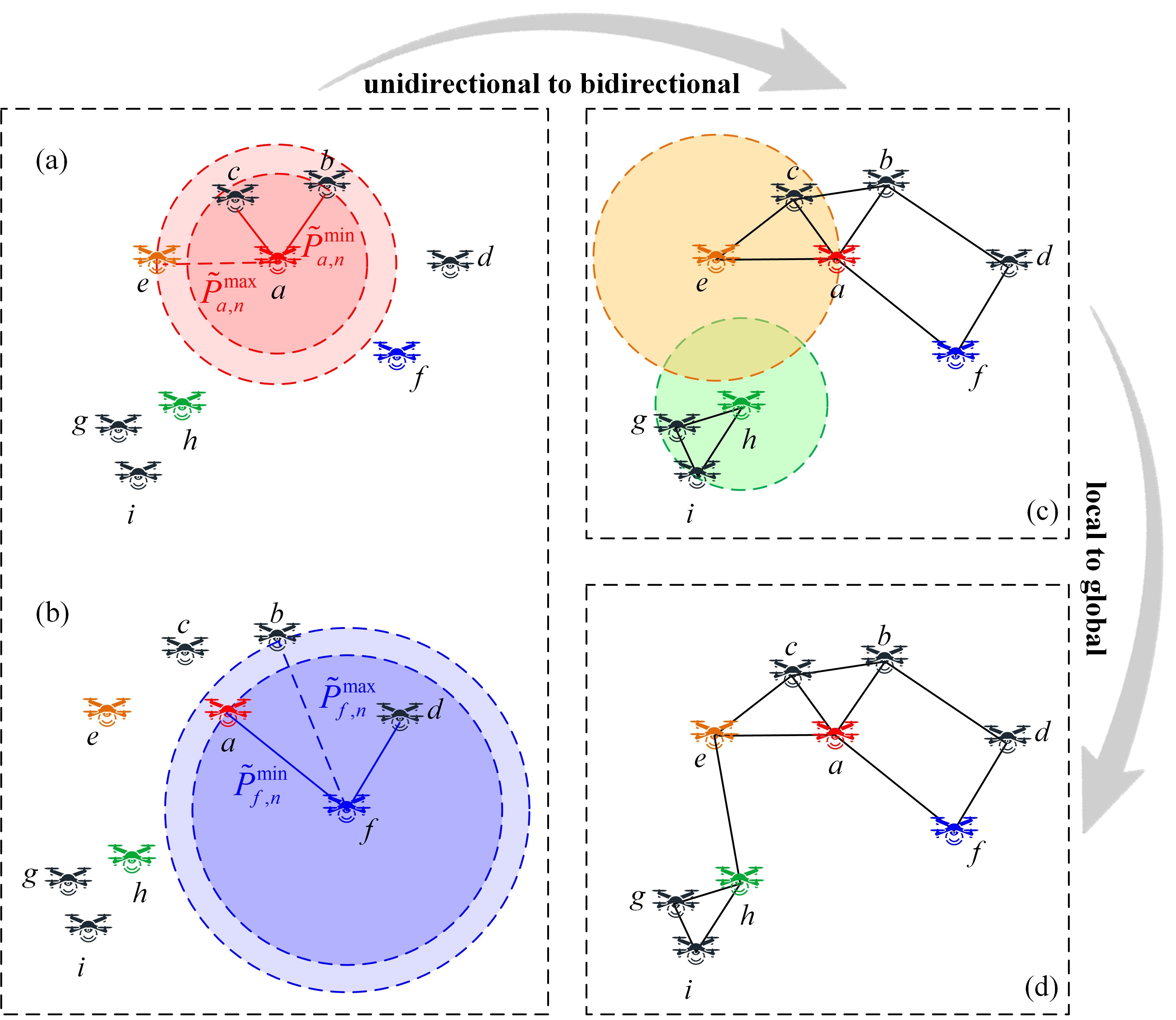}
\caption{Topology diagrams. Figures (a) and (b) indicate that the neighbors determined by each UAV are not reciprocal, figure (c) shows a bidirectional topology that satisfies the constraints on the number of neighbors but is not globally connected, and figure (d) shows a global topology.}
\label{topology_all}
\vspace{-1em}
\end{figure}

\subsubsection{Determine the maximum powers}\label{max_power}
The minimization of the objective value of problem (\ref{P3}) is approximately solved by minimizing the number of neighbors of each UAV, subject to the constraint of maintaining a minimum number of neighbors. The variable $\tilde P_{a,n}^{\rm max}$ is chosen to minimize the number of neighbors for each UAV. Therefore, the value of $\tilde P_{a,n}^{\rm max}$ is determined by the distance between UAV $a$ and its $(K_{\rm min}+1)$-th nearest neighbor, UAV $e$. In conjunction with the constraint (\ref{P3pmax}), the solution for $\tilde P_{a,n}^{\rm max}$ is
\begin{eqnarray}
    \tilde P_{a,n}^{\rm max} &=& \min\{\frac{\gamma d_{a,e,n}^2} {\mu_f},P_{a, \rm max}\} \label{Pmax1}
\end{eqnarray}

By restricting the transmit power of UAV $a$ to the interval $[\tilde P_{a,n}^{\rm min}, \tilde P_{a,n}^{\rm max})$, the neighbors of UAV $a$ can be determined, as shown in Fig.~\ref{topology_all}(a), where $K_{\rm min}=2$ is assumed. The neighbors of UAV $a$ are connected to it with solid lines. Note that UAV $e$ is not considered a neighbor of UAV $a$ because the interval of the transmit power is half-open.

\subsubsection{Establish bidirectional links}
Using (\ref{Pmin1}) and (\ref{Pmax1}), each UAV can determine its neighbors individually. In some cases, however, the neighbors determined by each UAV are not reciprocal. An example is shown in Fig. \ref{topology_all}(b). We can see from Fig. \ref{topology_all}(a) and Fig. \ref{topology_all}(b) that UAV $f$ is not a neighbor of UAV $a$ as $\tilde P_{a,n}^{\rm min} = \frac{\gamma d_{a,b,n}^2} {\mu_f}$ is used, while UAV $a$ is the neighbor of UAV $f$ as $\tilde P_{f,n}^{\rm min} = \frac{\gamma d_{a,f,n}^2} {\mu_f}$ is used and $d_{a,f,n}> d_{a,b,n}$. In such cases, the minimal transmit power of UAV $a$ is increased to be equal to $\tilde P_{f,n}^{\rm min}$. 
\textcolor{black}{Assume UAV $f$ is the $(K_{\rm min} + \kappa)$-th neighbor of UAV $a$. In this case, $\tilde{P}^{\rm max}_{a,n}$ is updated to the value required to cover its $(K_{\rm min} + \kappa +1)$-th neighbor, which is given by $\tilde{P}^{\rm max}_{a,n} = \frac{\gamma d_{a,d,n}^2} {\mu_f}$ in the example of Fig.~\ref{topology_all}}. 
The resultant topology with reciprocal neighbors is shown in Fig. \ref{topology_all}(c). Different from the traditional K-neigh algorithm \cite{Kneigh}, which abandons the one way link, a minimum number of neighbors is guaranteed in the proposed C-TOP algorithm by establishing a two-way link.

\subsubsection{Establish global topology}
It can be seen that the constraint in (\ref{P3_fullC}) is still not considered in the establishment of the topology, without which the resultant topology may consist of several isolated clusters, causing the network to be disconnected. As shown in Fig.~\ref{topology_all}(c), while the number of neighbors of each UAV satisfies the constraints in (\ref{P3kmin}), (\ref{P3kmax}) and (\ref{P3pmax}), the UAVs in the left cluster are not connected to those in the right cluster. The solution is to increase the transmit power of UAVs $e$ and $h$ using $d_{e,h}$ in (\ref{Pmin1}). The final global topology is shown in Fig.~\ref{topology_all}(d).



The optimal topology of the network $\mathcal{G}^\star = \{(\mathcal{A},{\bf C}_n^{\star}),n \in \mathcal{N}\}$ is obtained once the optimal values of $\tilde P_{a,n}^{\rm min}$ and $\tilde P_{a,n}^{\rm max}$ are determined. 
\textcolor{black}{The number of neighbors for each UAV in the optimized topology is guaranteed to exceed $K_{\rm min}$ for two reasons. First, during path planning, this constraint is ensured by the restriction in (\ref{P1kmin}). Second, in the topology optimization process, the condition in (\ref{Pmin1}) controls the number of neighbors while determining the lower bound of transmit powers.}
Finally, the constraint in (\ref{P3}) is verified. If the constraint is violated, i.e. $\tilde \delta \triangleq k_{a,n}^{\rm max} - K_{\rm min} > \delta$, then $(\tilde \delta - \delta)$ neighbors of node $a$ are removed from its list of neighbors. Otherwise, the optimal topology is used to optimize the transmit powers of the UAVs in the next subsection. 

\begin{algorithm}[!t]
\caption{Topology Optimization}\label{algorithm2_simple}
\KwIn{Optimal path set $\mathcal{J}^*$}
\KwOut{Optimal topology set $\mathcal{G}^*$}
$\bf{D} \leftarrow \mathcal{J}^*$

\For{$n=1$ to $N$}
    {Calculate $\tilde {\bf P}_{n}^{{\rm{min}}}$, $\tilde {\bf P}_{n}^{{\rm{max}}}$
    
    Determine ${\bf{C}}_n$
    
    \While{${\bf{C}}_n$ is not symmetric}
        {$a, b \leftarrow$ unidirectional links
        
        Update $[\tilde P_{a,n}^{{\rm{min}}},\tilde P_{a,n}^{{\rm{max}}})$, $[\tilde P_{b,n}^{{\rm{min}}}, \tilde P_{b,n}^{{\rm{max}}})$
        
        Update ${\bf{C}}_n$} 
    
    \While{${\bf{C}}_n$ is not globally connected}
        {$a, b \leftarrow$ the shortest unconnected link
        
        Update $[\tilde P_{a,n}^{{\rm{min}}},\tilde P_{a,n}^{{\rm{max}}})$, $[\tilde P_{b,n}^{{\rm{min}}}, \tilde P_{b,n}^{{\rm{max}}})$
        
        Repair unidirectional links
        
        Update ${\bf{C}}_n$}
        }

$\mathcal{G}^* \leftarrow (\mathcal{A},{\bf{C}}^*_n)$

\end{algorithm}

It is worth noting that the network topology optimized in this subsection remains unchanged even if the transmit power $p_{a,n}$ of the UAVs is determined in the next subsection. \textcolor{black}{The reason is that the network topology is controlled by the interval of the transmit powers shown in (\ref{Pmin1}) and (\ref{Pmax1}) 
throughout the entire process of topology optimization. This range ensures that the number of neighbors for each UAV remains fixed, even as the transmit power varies.
} 
Therefore, changing the transmit powers impacts the network throughput but does not affect the network topology. 
\textcolor{black}{Moreover, the ranges of transmit powers can be determined in polynomial time. From the pseudocode of topology optimization in Alg.~\ref{algorithm2_simple}, it can be seen that the computational complexity is mainly attributed to the node traversal during power control. The complexity for each node to update its power is $\mathcal{O}(A^2)$, where $A$ is the number of UAVs. This operation is repeated $A^2$ times over $N$ time slots. Consequently, the overall computational complexity can be expressed as $\mathcal{O}(A^4 N)$, and the space complexity is $\mathcal{O}(A^2 N)$, as space is used to store the power values and the node states for each time slot.}


\subsection{Throughput Maximization}
As illustrated in Sections \ref{sec-trajectory} and \ref{sec-topology}, the trajectories and topologies of the network have been optimized, which determines the optimal interval of the transmit powers of the UAVs. Therefore, in this subsection, the transmit power of each UAV is optimized to maximize the network throughput using the C-TOP algorithm. Consequently, the optimization problem in (\ref{P0}) is thus reduced to
\begin{subequations}\label{P4}
\allowdisplaybreaks
    \begin{align}
        \underset {\underset{a \in \mathcal{A}, n \in \mathcal{N}} {p_{a,n}}} \max & \; \sum_{n=1}^{N} \sum_{b=a+1}^{A} \sum_{a=1}^{A} c_{a,b,n}^{\star} R_{a,b,n}& \label{P4obj}\\
        {\rm s.t.}\quad & \tilde P_{a,n}^{{\rm min}} \leq p_{a,n} \leq \tilde P_{a,n}^{{\rm max}}, & a\in \mathcal{A}, n \in \mathcal{N} \label{P4pmin}\\
        &\sum_{n = 0}^N p_{a,n}  \leq E_{a, \rm max }, &a\in \mathcal{A} \hspace*{3.2em} \label{P4emax}\\
        & J_a \in \mathcal{J}^{\star}, & a\in \mathcal{A} \hspace*{3.2em} \label{P4J}
    \end{align}
\end{subequations}
where $c_{a,b,n}^{\star}$ is the element of the reachability matrix ${\bf C}_n^{\star}$ corresponding to the optimal topology. From (\ref{Rab}), it can be seen that the throughput of each link $R_{a,b,n}$ is a concave function of the transmit power $p_{a,n}$. Moreover, the summation in (\ref{P4obj}) is a linear operation, which preserves convexity. As a result, the maximization of the total network throughput in (\ref{P4obj}) is a convex problem. Furthermore, the constraints (\ref{P4pmin}) and (\ref{P4emax}) are linear. Consequently, problem (\ref{P4}) is convex and can be solved efficiently using the well-established interior point method in polynomial time, \textcolor{black}{which has a computation complexity of ${\mathcal O}( {{(AN)^{3.5}}})$ \cite{SOCP}, and a space complexity of ${\mathcal O}((AN)^2)$. The flow chart of C-TOP algorithm is shown in the green parts of Fig.~\ref{system_diagram}.
}

\textcolor{black}{\textit{Discussions}: From this section, it can be seen that the NP-hard problem in (\ref{P0}) is approximately solved by three sub-problems, (\ref{P1}), (\ref{P3}), and (\ref{P4}), sequentially. While decoupling the original problem into these sub-problems provides an efficient and scalable approach, it does not always guarantee the global optimum of the original joint optimization problem. This decoupling strategy works effectively when the coupling among topology, power, and trajectory optimization is weak, and the sub-problems are convex, enabling the sequential solving of sub-problems to achieve near-optimal results. 
Although the problem is divided into separate phases of path planning and topology optimization, these two phases are interdependent, as reflected by constraints (\ref{P1kmin}) and (\ref{P4pmin}). The constraint in (\ref{P1kmin}) directly restricts the number of neighbor nodes to maintain the desired network topology, while the constraint in (\ref{P4pmin}) limits the transmit power, which also impacts the number of neighbor nodes. This interdependency ensures that the decoupled approach remains aligned with the overarching network requirements. 
}

\textcolor{black}{In addition, the method developed in this paper addresses a CVRP-like problem with additional constraints on communication topologies and data throughput, making it particularly well-suited for wide-area environmental monitoring and data collection using multiple UAVs, especially in scenarios involving real-time data transmission and precise UAV positioning. Consequently, it has potential applications in other combinatorial optimization problems and robust control optimization problems for uncertain nonlinear multi-agent systems (MAS) with similar constraints \cite{MAS1}-\cite{ECBS}, extending its applicability beyond the specific context discussed in this study.}

\textcolor{black}{Despite its advantages, the proposed approach has certain limitations. In particular, its performance depends on accurate prior information, such as the DEM and the path loss model. Moreover, the interdependence between topology and path optimization is only partially addressed, which may limit the algorithm's effectiveness in dynamic environments. Addressing these challenges is part of our future work, which focuses on developing online multi-UAV task scheduling and topology self-healing strategies. These strategies are intended to enhance the system's adaptability to emergencies and real-time planning requirements in dynamic environments, thereby improving the completeness and robustness of the multi-UAV system.}



\section{Experimental Results and Analysis}
To validate the effectiveness of the proposed algorithm in this paper, we conduct both computer simulations and field experiments in this section. The simulation is carried out in a mountainous region located in Wanglang, Sichuan, China. In the field experiments, the target area is Wanghai Mountain, situated in Tianjin, China.

\subsection{Simulation Experiments}
\begin{figure}[!t]
\centering
\includegraphics[width=3in]{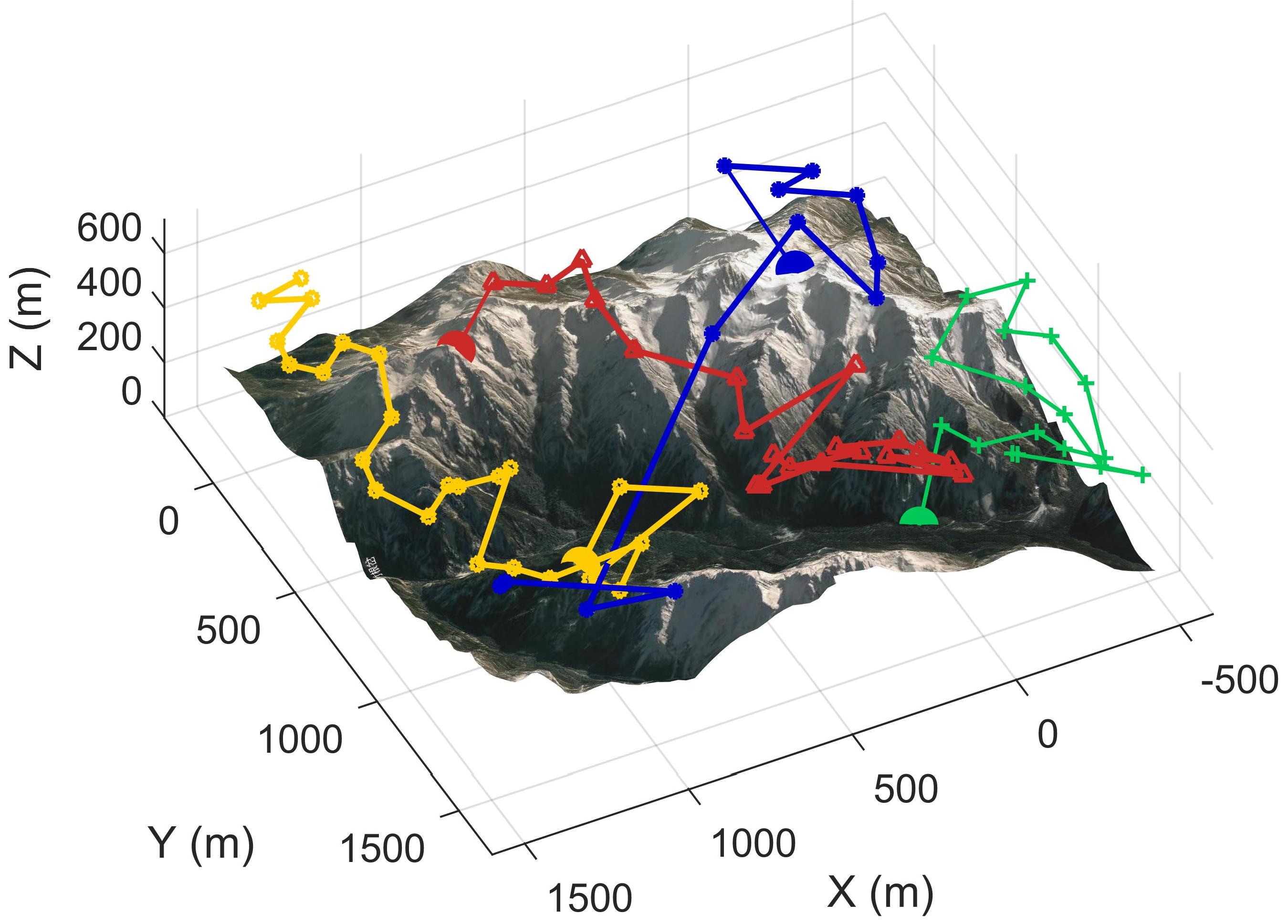}
\caption{Example paths of four UAVs. The solid circles represent four starting points.}
\label{path_Wanglang}
\vspace*{-1em}
\end{figure}

The 3D view of Wanglang Mountain is shown in Fig. \ref{path_Wanglang}. The area spans $2$~KM $\times$ $2$~KM, which can be visually covered by traversing $S=4$ starting points and $W=136$ waypoints. A total of $8$ \textcolor{black}{multirotor} UAVs are deployed to traverse all the waypoints within a time frame of $T_{a,\rm{max}} \in [400,450]$~s. The simulation uses $N = 150$ time slots, assuming a quasi-static topology and communication channels for each time-slot. The total path length is constrained $L_{\rm max} = 40$~KM, \textcolor{black}{which is determined by the specific requirements of the target applications, as well as the inherent hardware limitations of UAVs.}

In our experiments, each UAV is equipped with a battery capacity of $E_{a, \rm max} \in [150,200]$~J for data communication. The maximal allowed transmit power for the UAVs is $P_{a, \rm max} \in [27,30]$~dBm and the sensitivity threshold for the receive power is $\gamma = -70$~dBm. The channel bandwidth is $83.5$~MHz, and received noise power is $-110$~dBm. The minimal number of neighbors $K_{\rm min}$ is set to $2$, with the parameter $\delta$ set to 2.

Four of the resultant paths are shown in Fig. \ref{path_Wanglang}, marked with circle, triangle, rhombus, and star symbols, respectively. The remaining paths are not shown for clarity. The total length of the optimal path obtained using the proposed modified RL-TP is $30.83$~KM. \textcolor{black}{As seen from the figure, each waypoint is visited exactly once by a single UAV. However, some of the routes intersect, which is not explicitly avoided in our optimization. These intersections allow the UAVs to capture the target from multiple angles, which is beneficial since the cameras are required to be frontal to the target area, and their orientations may vary.}

\begin{figure}[!t]
\centering
\includegraphics[width=3in]{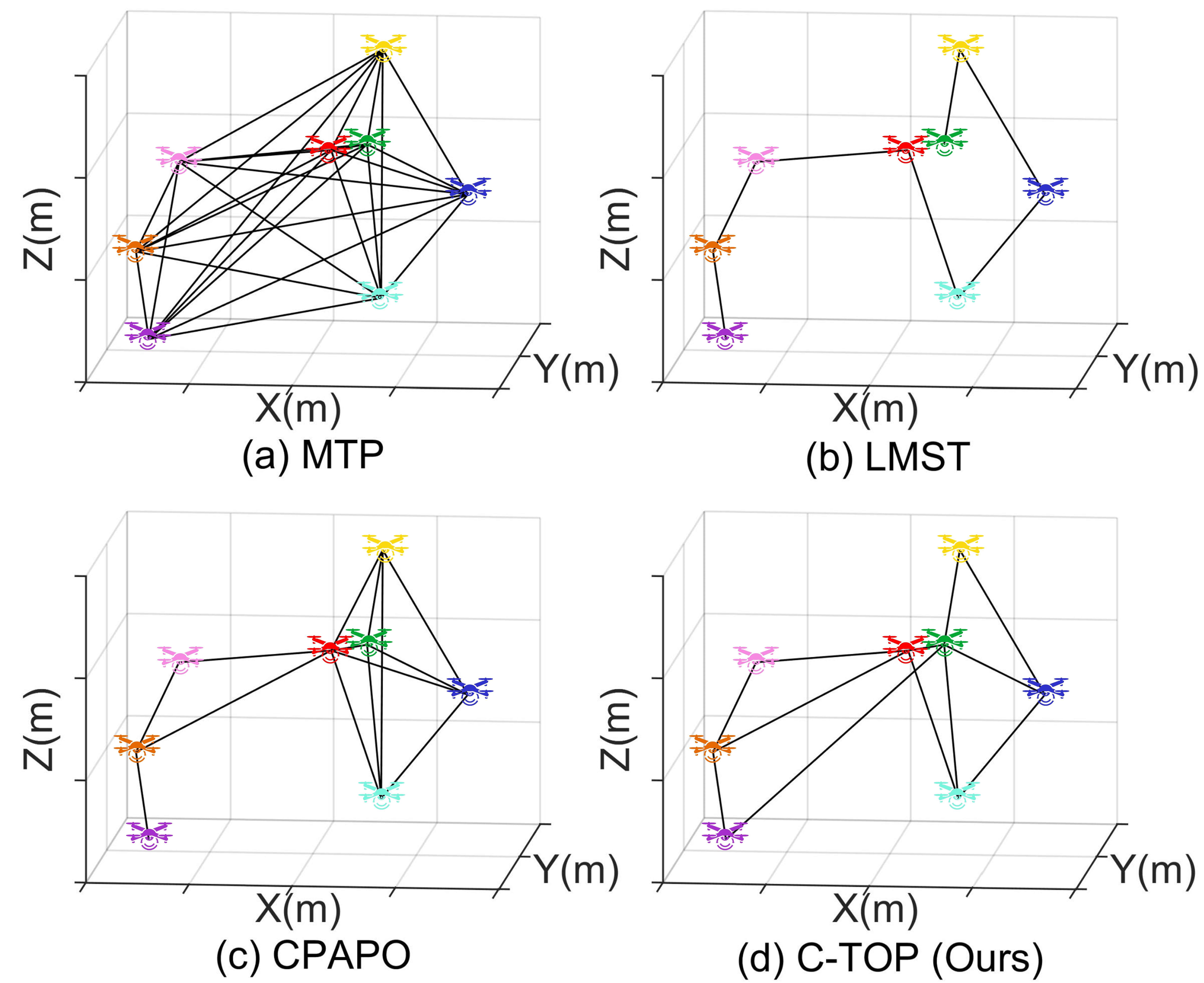}
\caption{Network topologies at time slot $n=45$. Figure (a)-(d) are the MTP, A-LMST, CPAPO and C-TOP (Ours) algorithms, respectively.}
\label{topo300_Wanglang}
\vspace*{-1em}
\end{figure}

The time-varying topology is optimized using the C-TOP algorithm in (\ref{P3}) as the UAVs travel along the planned path. Fig.~\ref{topo300_Wanglang}(d) illustrates an example of the network topology at the $45$-th time slot, where solid line segments represent valid data connections. As seen in the figure, the topology satisfies the constraints regarding the number of neighbors.


\begin{figure}[!t]
\centering
\includegraphics[width=3in]{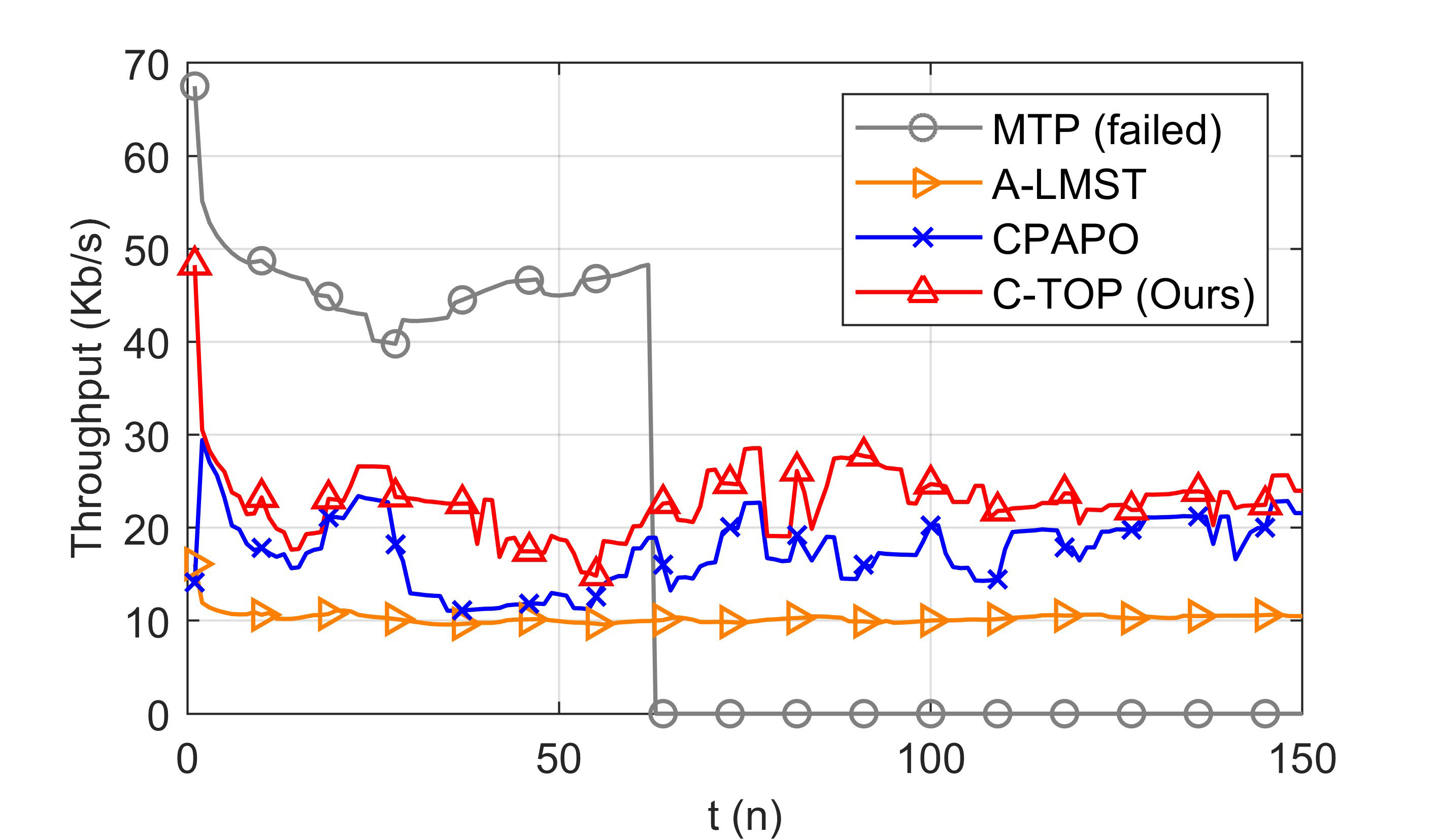}
\caption{Network throughput at different time slots. The throughput of the proposed C-TOP algorithm is significantly higher than that of the other algorithms. Although the MTP algorithm achieves the highest throughput in the first 60 time slots, it eventually depletes the energy supply before fully covering the target area.}
\label{throughput}
\vspace*{-1em}
\end{figure}


\begin{table*}[!t]
    \caption{\textbf{Comparison of simulation results}}
    \label{table_connectivity}
    \centering
    \begin{tabular}{cccccccc}
        \toprule
        Scene &Algorithms &\makecell[c]{Traverse all \\ the waypoints}  &$\xi$  &Throughput (Kb/s)  & Average Hop\\
        \midrule
        \multirow{4}{*}{\makecell[c]{$A=4$ \\ $N=50$ \\ 500M$\times$500M}}
        &MTP  &No               &94.96\%             &$1.90\times 10^3$     &$inf$\\
        &A-LMST &\textbf{Yes}      &81.72\%            &$6.94\times 10^2$     &1.64\\
        &CPAPO  &\textbf{Yes}      &89.74\%            &$9.50\times 10^2$     &1.32\\
        &\textbf{C-TOP (Ours)}  &\textbf{Yes}   &\textbf{97.20\%}   &\bm{$1.93\times 10^3$}     &\textbf{1.18}\\

        \midrule
        \multirow{4}{*}{\makecell[c]{$A=8$ \\ $N=150$ \\ 2000M$\times$2000M}}
        &MTP  &No             &39.69\%             &$8.28\times 10^3$     &$inf$  \\
        &A-LMST &\textbf{Yes}      &63.76\%            &$4.47\times 10^3$     &2.61  \\
        &CPAPO  &\textbf{Yes}      &77.22\%            &$7.65\times 10^3$     &1.85 \\
        &\textbf{C-TOP (Ours)}  &\textbf{Yes}  &\textbf{88.21\%}   &\bm{$9.89\times 10^3$}     &\textbf{1.59}  \\

        \midrule
        \multirow{4}{*}{\makecell[c]{$A=8$ \\ $N=150$ \\ 2000M$\times$2000M \\ Gaussian distribution}}
        &MTP  &No              &37.80\%             &$6.36\times 10^3$     &$inf$ \\
        &A-LMST &\textbf{Yes}    &72.89\%            &$4.26\times 10^3$     &2.64  \\
        &CPAPO  &\textbf{Yes}    &93.03\%            &$9.07\times 10^3$     &1.73 \\
        &\textbf{C-TOP (Ours)}  &\textbf{Yes} &\textbf{95.36\%}   &\bm{$1.07\times 10^4$}     &\textbf{1.61}  \\
        
        \bottomrule
    \end{tabular}
\end{table*}

\textcolor{black}{The performance of the proposed algorithm is compared with three other algorithms under identical system parameter configurations, such as target area, number, type, and intrinsic characteristics of UAVs.} The three algorithms compared in this study are maximal-transmit-power (MTP) \textcolor{black}{\cite{MTP}}, adaptive local MST (A-LMST) \cite{ALMST}, and cyclic pruning-assisted power optimization (CPAPO) \cite{CDS_throughput}. In MTP, each UAV transmits signals at the maximum allowed power, $P_{a, \rm max}$. The topologies generated by these algorithms are shown in Fig. \ref{topo300_Wanglang}(a), Fig. \ref{topo300_Wanglang}(b), and Fig. \ref{topo300_Wanglang}(c), respectively. We observe that MTP results in a significantly higher number of neighbors compared to the proposed method, leading to low spatial reuse in the network. Both A-LMST and CPAPO produce topologies where the minimal number of neighbors is below $K_{\rm min}$, indicating reduced connectivity compared to the proposed C-TOP-based approach. As shown in Fig. \ref{topo300_Wanglang}(b), and Fig. \ref{topo300_Wanglang}(c), some UAVs in these topologies only establish a single neighbor. If this neighbor fails (e.g., due to a UAV malfunction), the networks designed by A-LMST and CPAPO become disconnected. Define
\begin{equation}
    \xi  \buildrel \Delta \over = \frac{1}{{A\left( {N - {n_{{\rm{DC}}}} + 1} \right)}}\sum\limits_{a = 1}^A {\sum\limits_{n = {n_{{\rm{DC}}}}}^N {\frac{{A{C_a}(n)}}{{A - 1}}} }
\end{equation}
as the connectivity rate of network, where $n_{\rm DC}$ is the time slot at which one of the UAVs is disconnected, and $AC_a(n)$ is the number of UAVs still connected in the $n$-th time slot after UAV $a$ is disconnected. In this example, each UAV is sequentially disconnected at $n_{\rm DC} = 45$ to evaluate the connectivity and robustness of the network. The values of $\xi$ for the four algorithms are presented in Table \ref{table_connectivity}. To provide a more comprehensive performance comparison, we also evaluate the algorithms in a $500$~M $\times$ $500$~M scenario with a Gaussian distribution of waypoints. \textcolor{black}{The results indicate that the network optimized using the proposed C-TOP algorithm achieves higher connectivity than the other three algorithms.} Consequently, the C-TOP-based network demonstrates superior robustness compared to the others. 
\textcolor{black}{
This is because the constraint on the minimum number of neighbors is explicitly included in our optimization problem, as shown in (\ref{P3kmin}). This constraint ensures that each UAV has at least $K_{\rm min}$ neighbors, so even if one neighbor is disconnected, the UAV remains connected to at least $(K_{\rm min} - 1)$ other UAVs. In contrast, the A-LMST and CPAPO-based algorithms prioritize minimizing network connectivity costs, leading to UAVs operating at the minimum achievable power. As a result, some UAVs may end up with only one neighbor. If that neighbor is disconnected, the network's global connectivity is lost, which limits their robustness. Moreover, the low connectivity rate of the MTP-based algorithm is due to the UAVs running out of energy before all the waypoints are traversed, leading to a loss of connectivity in the remaining time slots.
}

The total data throughput of the network is presented in Table \ref{table_connectivity} and the throughput at each time slot is shown in Fig. \ref{throughput}. It is evident that both the total data throughput and the throughput at each time slot for the proposed network outperform those of the MTP, A-LMST and CPAPO-based networks. Table \ref{table_connectivity} also presents the average number of hops for the network with different algorithms. It can be seen that the C-TOP-based network has the lowest hop count across all scenarios, indicating that the C-TOP algorithm also achieves optimal performance in terms of the network's average delay.


\subsection{Field Experiments}
\begin{figure}[!t]
\centering
\includegraphics[height=2.0in]{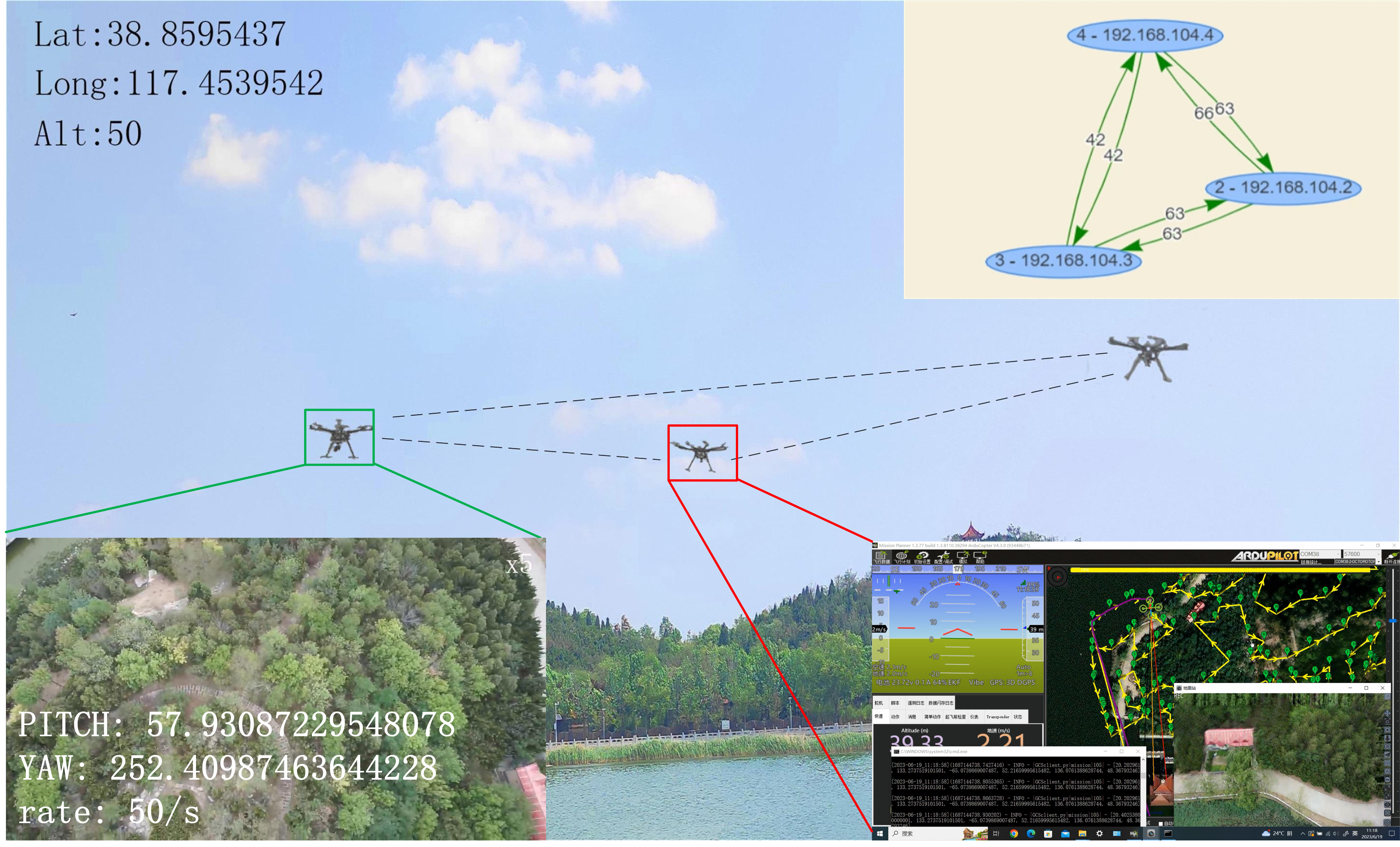}
\caption{Field experiment over Wanghai Mountain, Tianjin, China. The global real-time topology, the returned image and posture data of the green UAV, and the path monitoring screenshot of the red UAV are also shown in the figure.}
\label{fig_Wanghai}
\end{figure}

\begin{figure}[!t]
\centering
\includegraphics[height=2.0in]{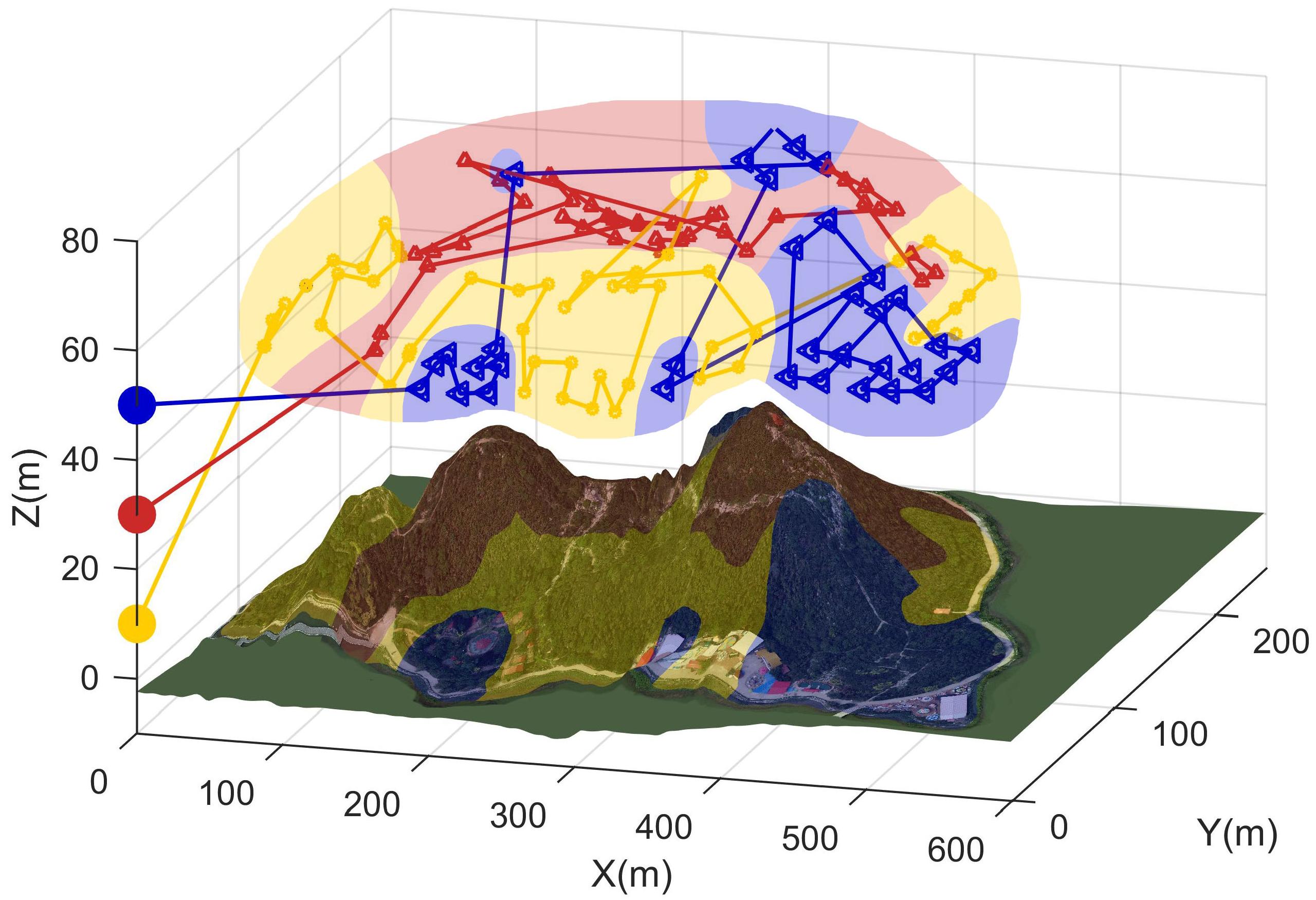}
\caption{Paths of UAVs for Wanghai Mountain.}
\label{path_Wanghai}
\vspace*{-1.em}
\end{figure}

The 3D view and overlaid paths of the Wanghai Mountain are shown in Fig. \ref{fig_Wanghai} and Fig. \ref{path_Wanghai}. The mountain features topographic relief and irregular regional boundaries, covering an area of $600$~M $\times$ $300$~M. In the experiment, $3$ \textcolor{black}{octocopter} UAVs are used for coverage and surveillance over $N=150$ time slots. There are $S=3$ starting points and $W=123$ waypoints. Three identical UAVs are used in the experiments for simplicity, although the proposed algorithm is also applicable to the scenarios using heterogeneous UAVs. The UAVs used in the experiments are DJI S1000+, equipped with the NexFi MFS series \emph{ad hoc} data communication module, 3B+ Raspberry PI and flight control of CUAV x7+. In addition, Mission Planner is used as the ground console.

\begin{table}[!t]
    \caption{\textbf{Comparison of field experimental results}}
    \label{table_PE}
    \centering
    \begin{tabular}{cccc}
        \toprule
        Algorithms &\makecell[c]{Traverse all \\ the waypoints} &$\xi$  &Throughput (Kb/s)\\
        \midrule
        MTP  &No               &63.61\%           &$5.39\times 10^3$\\
        A-LMST &\textbf{Yes}      &74.52\%            &$4.49\times 10^3$\\
        CPAPO  &\textbf{Yes}      &86.00\%            &$6.21\times 10^3$\\
        \textbf{C-TOP (Ours)}  &\textbf{Yes}  &\textbf{91.39\%}   &\bm{$7.71\times 10^3$}\\
        \bottomrule
    \end{tabular}
\end{table}

Table \ref{table_PE} shows a performance comparison between the proposed algorithm and three other algorithms \textcolor{black}{under identical hardware conditions}. When a UAV disconnects at $n_{DC}=45$, C-TOP improves the connectivity rate by at least $7.22\%$ compared to the other algorithms. Additionally, the maximum throughput of C-TOP is over 1.43 times greater than that of the other algorithms. The results are consistent with the simulation results. 

\vspace{-0.5em}
\section{Conclusions}
This paper presents a novel approach for the joint optimization of dynamic topology and path planning of multiple UAVs, which optimizes the data throughput of a time-varying FANET while planning the trajectories of multirotor UAVs. The optimization considers constraints on network topology, UAV trajectory length, and transmit power limitations. The resultant NP-hard problem is approximately solved using a three-step approach. Experimental results demonstrate that the proposed convex-based topology optimization (C-TOP) algorithm significantly outperforms adaptive local minimum spanning tree (A-LMST) and cyclic pruning-assisted power optimization (CPAPO) in terms of data throughput and network topology. Moreover, the proposed joint FANET topology optimization (JFTO) framework shows promising potential for wide-area monitoring scenarios.

\textcolor{black}{However, this paper does not address the system's robustness against parameter errors or consider dynamic environmental conditions. Therefore, our future work will focus on developing online multi-UAV task scheduling and topology self-healing strategies to improve adaptability and robustness in dynamic environments.}


\begin{thebibliography}{99}
\bibliographystyle{IEEEtran}

\bibitem{DP}
J.~Ren and X.~Huang, ``Dynamic programming inspired global optimal path planning for mobile robots,'' in \emph{2021 IEEE 4th Int. Conf. Inf. Syst. Comput. Aided Educ. (ICISCAE)}, pp. 461--465, 2021.

\bibitem{2022DCOC}
J.~Liu, ``An improved genetic algorithm for rapid UAV path planning,'' in \emph{J. Phys. Conf. Ser.}, vol. 2216, no.~1. IOP Publishing, p. 012035, 2022.

\bibitem{greedy}
\textcolor{black}{S.~Han, K.~Zhu, M.~Zhou, and X.~Liu, ``Joint deployment optimization and flight trajectory planning for UAV assisted IoT data collection: A bilevel optimization approach,'' \emph{IEEE Trans. Intell. Transp. Syst.}, vol.~23, no.~11, pp. 21\,492--21\,504, 2022.}

\bibitem{GA1}
\textcolor{black}{H.~Xie, D.~Zhang, X.~Hu, M.~Zhou, and Z.~Cao, ``Autonomous multi-robot navigation and cooperative mapping in partially unknown environments,'' \emph{IEEE Trans. Instrum. Meas.}, 2023.}

\bibitem{CVRP}
Y.-H. Jia, Y.~Mei, and M.~Zhang, ``A bilevel ant colony optimization algorithm for capacitated electric vehicle routing problem,'' \emph{IEEE Trans. Cybern.}, vol.~52, no.~10, pp. 10\,855--10\,868, 2022.

\bibitem{PSO}
\textcolor{black}{R.~Lai, B.~Zhang, G.~Gong, H.~Yuan, J.~Yang, J.~Zhang, and M.~Zhou, ``Energy-efficient scheduling in UAV-assisted hierarchical wireless sensor networks,'' \emph{IEEE Internet Things J.}, 2024.}

\bibitem{cvxpp}
H.~Zhou, Z.~Cai, J.~Zhao, and Y.~Wang, ``Rho-based convex optimization method applied to cooperative trajectory planning for multiple UAVs,'' in \emph{11th Asian Control Conf.}, pp. 1572--1577, 2017.

\bibitem{Attention}
\textcolor{black}{W.~Kool, H.~V. Hoof, and M.~Welling, ``Attention, learn to solve routing problems!'' in \emph{Int. Conf. Learn. Representations}, 2019.}


%

\bibitem{MST}
S.~Diaz and D.~Mendez, ``Dynamic minimum spanning tree construction and maintenance for wireless sensor networks,'' \emph{Rev. Facultad Ing. Universidad Antioquia}, no.~93, pp. 57--69, 2019.


\bibitem{RNG}
N.~Hassan, M.~Bakhouya, M.~Rachik, and J.~Gaber, ``Using energy efficient relative neighborhood graph for AODV routing protocol in MANETs,'' \emph{Int. J. Sci. Eng. Res.}, 2015.

\bibitem{Yao}
M.~Kadivar, ``An adaptive Yao-based topology control algorithm for wireless ad-hoc networks,'' in \emph{10th Int. Conf. Comput. Knowl. Eng.}, pp. 457--462, 2020.

\bibitem{CBTC}
L.~Wang, W.~Zhao, Y.~Li, Y.~Qu, Z.~Liu, and Q.~Chen, ``Sleep-supported and cone-based topology control method for wireless sensor networks,'' in \emph{IEEE Int. Conf. Networking Sens. Control}, pp. 1445--1448, 2008.

\bibitem{FTTC}
J.~Guo, J.~Cao, Y.~Ren, C.~Jiang, X.~Liu, ``Distributed fault-tolerant topology control in cooperative wireless ad hoc networks,'' \emph{IEEE Trans. Parallel Distrib. Syst.}, 2015.

\bibitem{Mobility1}
M.~Tan, L.~Fang, Y.~Wu, B.~Zhang, B.~Chang, P.~Holme, and J.~Zhao, ``A fault-tolerant small world topology control model in ad hoc networks for search and rescue,'' \emph{Phys. Lett. A}, vol. 382, no.~7, pp. 467--476, 2018.


\bibitem{Mobility3}
M.~M. Alam and S.~Moh, ``Joint topology control and routing in a UAV swarm for crowd surveillance,'' \emph{J. Net. Comput. Appl.}, vol. 204, 2022.



\bibitem{FT2}
H.~Ouchitachen, A.~Hair, and N.~Idrissi, ``Improved multi-objective weighted clustering algorithm in wireless sensor network,'' \emph{Egypt. Inf. J.}, vol.~18, no.~1, pp. 45--54, 2017.

\bibitem{Mobility4}
X.~Fu and X.~Gao, ``Multi-UAVs cooperative control in communication relay,'' in \emph{2016 IEEE Int. Conf. Signal Process. Commun. Comput. (ICSPCC)}. IEEE, pp. 1--5, 2016.

\bibitem{JO1}
K.~Wang, K.~Sun, J.~Xu, P.~Liu, Y.~Cong, and X.~Wang, ``Throughput maximization: Joint trajectory planning, power and time allocation in UAV network,'' in \emph{2023 IEEE Int. Conf. Unmanned Syst. (ICUS)}. IEEE, pp. 552--558, 2023.

\bibitem{JO2}
Y.~Zhang, M.~Pan, Q.~Han, W.~Long, and S.~Yang, ``Joint power allocation and route planning scheme for multitarget tracking in airborne radar network under multiuncertainty,'' \emph{IEEE Sens. J.}, vol.~23, no.~7, pp. 7705--7718, 2023.

\bibitem{JO3}
Y.~Wu, B.~Zhang, S.~Yang, X.~Yi, and X.~Yang, ``Energy-efficient joint communication-motion planning for relay-assisted wireless robot surveillance,'' in \emph{IEEE INFOCOM 2017-IEEE Conf. Comput. Commun.}. IEEE, pp. 1--9, 2017.

\bibitem{JO4}
H.~Wang, G.~Ding, J.~Chen, Y.~Zou, and F.~Gao, ``UAV anti-jamming communications with power and mobility control,'' \emph{IEEE Trans. Wireless Commun.}, 2022.

\bibitem{JO5}
M.~Huang, Y.~Chen, and X.~Tao, ``Proactive eavesdropping in UAV systems via trajectory planning and power optimization,'' in \emph{2021 IEEE Wireless Commun. Networking Conf. Workshops (WCNCW)}. IEEE, pp. 1--6, 2021.

\bibitem{2022DEM}
H.~Wang, S.~Zhang, X.~Zhang, X.~Zhang, and J.~Liu, ``Near-optimal 3-D visual coverage for quadrotor unmanned aerial vehicles under photogrammetric constraints,'' \emph{IEEE Trans. Ind. Electron.}, vol.~69, no.~2, pp. 1694--1704, 2022.


\bibitem{C_matrix}
\textcolor{black}{P.~Mittal, S.~Shah, and A.~Agarwal, ``Power-efficient joint link selection and multihop routing for throughput maximization in UAV assisted FANETs,'' in \emph{2022 IEEE 33rd Annu. Int. Symp. Pers. Indoor Mobile Radio Commun. (PIMRC)}.\relax IEEE, pp. 1282--1287, 2022.}

\bibitem{threshold}
\textcolor{black}{V.~Van~Khoa and T.~Shigeru, ``Topology control for wireless sensor network in landslide monitoring,'' in \emph{2019 58th Annu. Conf. Soc. Instrum. Control Eng. Japan (SICE)}. IEEE, pp. 394--399, 2019.}

\bibitem{path_loss}
\textcolor{black}{R.~E. De~Moraes, Y.~S. Silva, F.~N. Martins, J.~A. Silva, and H.~R. Rocha, ``Joint interference and power minimization for fault-tolerant topology in sensor networks,'' \emph{IEEE Access}, 2024.}

\bibitem{node_degree}
\textcolor{black}{T.-V.~T. Duong, V.~M. Ngo \emph{et~al.}, ``TFACR: A novel topology control algorithm for improving 5G-based MANET performance by flexibly adjusting the coverage radius,'' \emph{IEEE Access}, 2023.}

\bibitem{CDS_throughput}
X.~Qi, P.~Yuan, Q.~Zhang, and Z.~Yang, ``CDS-based topology control in FANETs via power and position optimization,'' \emph{IEEE Wireless Commun. Lett.}, vol.~9, no.~12, pp. 2015--2019, 2020.


\bibitem{space_reuse}
\textcolor{black}{J.~Ma, S.~Zhang, H.~Li, N.~Zhao, and V.~C. Leung, ``Interference-alignment and soft-space-reuse based cooperative transmission for multi-cell massive MIMO networks,'' \emph{IEEE Trans. Wireless Commun.}, vol.~17, no.~3, pp. 1907--1922, 2018.}


\newpage
\bibitem{reachable}
\textcolor{black}{H.~Wu and G.~Liu, ``The relationships between topologies and generalized rough sets,'' \emph{Int. J. Approximate Reasoning}, vol. 119, pp. 313--324, 2020.}



\bibitem{CVRPsolution}
L.~Feng, Y.~Huang, L.~Zhou, J.~Zhong, A.~Gupta, K.~Tang, and K.~C. Tan, ``Explicit evolutionary multitasking for combinatorial optimization: A case study on capacitated vehicle routing problem,'' \emph{IEEE Trans. Cybern.}, vol.~51, no.~6, pp. 3143--3156, 2021.

\bibitem{RLTP}
H.~Wang, S.~Song, Q.~Guo, D.~Xu, X.~Zhang, and P.~Wang, ``Cooperative motion planning for persistent 3D visual coverage with multiple quadrotor UAVs,'' \emph{IEEE Trans. Autom. Sci. Eng.}, pp. 1--10, 2023.

\bibitem{BN}
\textcolor{black}{A.~Nigam, R.~K. Sanodiya, P.~Joshi \emph{et~al.}, ``Generalized visual path following on jetbot using normalization with reinforcement learning,'' in \emph{2024 IEEE Int. Conf. Adv. Rob. Its Social Impacts (ARSO)}. \relax IEEE, pp. 247--252, 2024.}

\bibitem{FF}
\textcolor{black}{J.~Chen, Z.~Xiao, H.~Xing, P.~Dai, S.~Luo, and M.~A. Iqbal, ``STDPG: A spatio-temporal deterministic policy gradient agent for dynamic routing in SDN,'' in \emph{ICC 2020-2020 IEEE Int. Conf. Commun. (ICC)}. \relax IEEE, pp. 1--6, 2020.}


\bibitem{Kneigh}
D.~M. Blough, M.~Leoncini, G.~Resta, and P.~Santi, ``The k-neighbours protocol for symmetric topology control in ad hoc networks,'' \emph{Tech. Rep. IIT-TR-23/2002, Istituto di Informatica e Telematica}, 2002.

\bibitem{SOCP}
M.~S. Lobo, L.~Vandenberghe, S.~Boyd, and H.~Lebret, ``Applications of second-order cone programming,'' \emph{Linear Algebra Appl.}, vol. 284, no. 1-3, pp. 193--228, 1998.

\bibitem{MAS1}
\textcolor{black}{H.~Ren, H.~Ma, H.~Li, and Z.~Wang, ``Adaptive fixed-time control of nonlinear mass with actuator faults,'' \emph{IEEE/CAA J. Autom. Sin.}, vol.~10, no.~5, pp. 1252--1262, 2023.}

\bibitem{MAS2}
\textcolor{black}{H.~Ren, Z.~Liu, H.~Liang, and H.~Li, ``Pinning-based neural control for multiagent systems with self-regulation intermediate event-triggered method,'' \emph{IEEE Trans. Neural Networks Learn. Syst.}, 2024.}

\bibitem{ECBS}
\textcolor{black}{S.~Feng, L.~Zeng, J.~Liu, Y.~Yang, and W.~Song, ``Multi-UAVs collaborative path planning in the cramped environment,'' \emph{IEEE/CAA J. Autom. Sin.}, vol.~11, no.~2, pp. 529--538, 2024.}

\bibitem{MTP}
\textcolor{black}{R.~Kirichek, V.~Vishnevsky, A.~Koucheryavy \emph{et~al.}, ``Analytic model of a mesh topology based on LoRa technology,'' in \emph{2020 22nd Int. Conf. Adv. Commun. Technol. (ICACT)}. IEEE, pp. 251--255, 2020.}

\bibitem{ALMST}
L.~B. Abiuzi, C.~D.~A. Cesar, and C.~H. Ribeiro, ``A-LMST: An adaptive LMST local topology control algorithm for mobile ad hoc networks,'' in \emph{2016 IEEE 41st Conf. on Local Comput. Networks (LCN)}, pp. 168--171, 2016.



\end{thebibliography}
\end{document}